\documentclass[preprint,12pt]{elsarticle}

\usepackage{amsmath,amsfonts,amsthm,amssymb,paralist,subfigure,graphicx,amsbsy,float,epsfig,color}
\usepackage{cuted,mathtools,lipsum}

\usepackage[ruled,vlined,linesnumbered]{algorithm2e}
\usepackage{stmaryrd,url}

\usepackage{amsthm}
\newtheorem{lem}{Lemma}
\newtheorem{ass}{Assumption}
\newtheorem{theorem}{Theorem}
\newtheorem{defn}{Definition}
\newtheorem{rem}{Remark}

\def\mb{\mathbf}

\def\mc{\mathcal}

\def\mb{\mathbf}

\def\mc{\mathcal}

\journal{Nonlinear Analysis: Hybrid Systems}

\begin{document}

\begin{frontmatter}



\title{Low-Bit Distributed Constrained Optimization with Application to Optimal Scheduling and Allocation
}


\author[Sem]{Mohammadreza Doostmohammadian}
\affiliation[Sem]{Mechatronics Group, Faculty of Mechanical Engineering, Semnan University, Semnan, Iran, doost@semnan.ac.ir.}

\author[aa]{ Sergio Pequito}
\affiliation[aa]{Department of Electrical and Computer Engineering and Institute for Systems and Robotics, Instituto Superior Tecnico, University of Lisbon, Portugal, sergio.pequito@tecnico.ulisboa.pt.}

\begin{abstract}
This paper proposes a distributed resource allocation algorithm in which each agent communicates only the sign of its local cost gradient---a single bit per directed link per iteration. The algorithm is based on Laplacian-gradient dynamics and preserves the equality (demand) constraint at \emph{every} iterate, guaranteeing anytime feasibility irrespective of termination time. We make three technical contributions.

First, we provide a complete Lyapunov-based convergence analysis for the fixed-step-size regime: the iterates are shown to enter, in at most $O(1/\varepsilon^2)$ iterations, a neighborhood of the optimizer whose radius scales as $O(\tau)$ in the decision variable and $O(\tau^2)$ in the cost residual, where $\tau$ is the step-size. Crucially, the step-size bound is a \emph{time-invariant} network constant (depending only on the graph Laplacian spectrum and the cost curvature bounds), so no adaptive tuning is required.

Second, we characterize the invariant set under $B$-connected, weight-balanced digraphs---strictly weaker than the all-time connectivity and doubly-stochastic weights assumed in most prior work---and show that any odd, sign-preserving nonlinearity (not only the signum) yields the same invariant-set and descent structure, with the single-bit protocol being the extreme communication-minimizing instance of a broader nonlinear class.

Third, we address local box constraints via smooth penalty relaxation and quantify explicitly how the penalty weight interacts with the step-size bound and the optimality gap.

Simulations on economic dispatch (1\,000 generators) and CPU scheduling (500 servers) demonstrate that the single-bit protocol achieves $5$--$60\times$ reduction in total communication cost (bits $\times$ iterations to a target accuracy) relative to multi-bit baselines, at the price of a tunable and explicitly bounded optimality gap. 
A diminishing step-size schedule is shown empirically to eliminate this gap; its formal convergence analysis is identified as an open problem.
\end{abstract}

\begin{graphicalabstract}
	\includegraphics{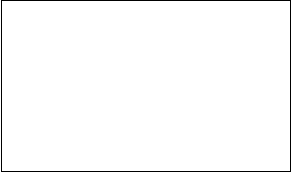}
\end{graphicalabstract}

\begin{highlights}
	\item Defining formal mathematical optimization framework for distributed resource allocation over multi-agent systems 
	\item Proposing a distributed setup for optimal convergence and analyzing its feasibility, convergence, and network connectivity
	\item Proposing an algorithm with low bit data exchange over the network to reduce the communication and networking traffic 
\end{highlights}


\end{frontmatter}

\section{Introduction} \label{sec_intro}
Distributed sum-preserving optimization underpins resource allocation in CPU scheduling~\cite{rikos2021optimal,eaai}, autonomous vehicle battery charge scheduling~\cite{falsone2020tracking}, cyber-physical energy systems~\cite{chen2020distributed,guan2025port}, economic dispatch~\cite{cherukuri2015tcns,ma2019resource}, and automatic generation control~\cite{dsp}. A common thread in the recent literature is the assumption of \emph{infinite-precision} data exchange. For instance, projection-based methods~\cite{SUN2023109266}, drift-plus-penalty algorithms~\cite{wang2023distributed}, privacy-preserving gradient schemes~\cite{an2024privacy}, event-triggered accelerated algorithms~\cite{luo2024pda}, second-order disturbance-rejection methods~\cite{shi2023distributed,10422834}, and random coordinate descent for open systems~\cite{10508935}. Nonetheless, this demands high-bandwidth, low-latency links that may be unavailable in large-scale or unreliable networks.

Low-bit communication has been studied for diffusion strategies~\cite{sayin2013single} and consensus~\cite{taes2020finite,Scientia2011}, and sign-of-relative-state methods have been applied to unconstrained optimization~\cite{zhang2018distributed,cao2021decentralized}. Quantized coordination for quadratic objectives appears in~\cite{rikos2021optimal}. None of these works addresses \emph{constrained} resource allocation under a single-bit regime. On the application side, distributed economic dispatch has received considerable attention~\cite{kar_edp,zhou2024fixed,ding2021differentially,wang2022byzantine,alizadeh2012demand,chen2016distributed,kar2012distributed,casagrande2022resilient,sarafraz2022uncertain}; all existing methods require multi-bit exchange and, with the exception of Laplacian-gradient (LG) schemes, typically rely on all-time network connectivity and doubly-stochastic weights.

{A related line of work addresses communication efficiency in federated learning through zeroth-order (ZO-FL) optimization. In \cite{fang2022communication}, the authors propose FedZO, a derivative-free FL method that constructs stochastic gradient estimators from function evaluations, thereby enabling optimization when explicit gradient information is unavailable; an over-the-air computation variant is further developed to reduce the communication burden in wireless FL. More recently, \cite{li2025achieving2} proposes DeComFL, which exploits the decomposition of a ZO gradient estimator into a scalar gradient magnitude and a shared perturbation direction to reduce the per-round communication from $\mathcal{O}(d)$ to $\mathcal{O}(1)$, independent of the dimension $d$ of the optimization variable. This direction is pushed further in \cite{li2025achieving}, where only the signs of the ZO gradient scalars, together with the information needed to reconstruct the common perturbation directions, are communicated, yielding an extremely low communication footprint while retaining applicability to high-dimensional model parameters.}

Reducing the per-iteration communication payload is therefore a pressing practical concern: lower bandwidth consumption enables deployment over resource-constrained or unreliable links, simplifies protocol design by replacing floating-point encoding/decoding with a single binary flag, and accelerates coordination in latency-sensitive applications such as real-time CPU scheduling and grid-integrated energy dispatch. Henceforth, this paper addresses the following question: \emph{can distributed resource allocation be solved when each agent communicates only a single bit per iteration?} We answer affirmatively by proposing a single-bit LG dynamics in which every agent broadcasts solely the sign of its local gradient {(with a one-time multi-bit feasibility initialization)}. Specifically, we make the following main contributions relative to the state-of-the-art:
\begin{enumerate}
	\item \textbf{Single-bit communication.}
	The proposed algorithm requires each agent to transmit only $\{+1,-1\}$ per link per iteration. {This is specifically for scalar-valued resource variables as in many distributed resource allocation formulations, e.g., economic dispatch, load balancing, bandwidth allocation, charging-rate allocation, and proportional sharing problems, where each agent contributes one scalar amount to satisfy a global demand constraint.}
	This is, to the best of our knowledge, the first distributed resource allocation scheme with provable convergence under one-bit data exchange {(with a one-time multi-bit feasibility initialization)}, and it yields lower per-iteration computational cost than existing methods (verified empirically in Section~\ref{sec_sim}).

	\item \textbf{Anytime feasibility.}
	The equality constraint  is preserved at \emph{every} iteration, not merely in the limit. This is a strict improvement over ADMM-based approaches~\cite{dominguez2012decentralized,li2017distributed,falsone2020tracking}, which achieve feasibility only asymptotically and therefore cannot guarantee constraint satisfaction upon early termination.

	\item \textbf{Weak connectivity and weight-balance.}
	Convergence is established under \emph{uniform} connectivity of the time-varying directed graph, together with a weight-balance condition. Both requirements are strictly weaker than the all-time connectivity and doubly-stochastic weights assumed in~\cite{dominguez2012decentralized,li2017distributed,doan2017scl,boyd2006optimal,zhou2024fixed,kar_edp}.

	\item \textbf{General nonlinear analysis.}
	{The Lyapunov-based convergence proof applies to \emph{any} odd, sign-preserving nonlinearity (i.e., not only the signum function),} and accommodates non-quadratic objectives with smooth penalty terms for box-constraint handling.
\end{enumerate}

\paragraph{Outline}
Section~\ref{sec_def} formulates the problem and collects auxiliary lemmas. The single-bit algorithm is presented in Section~\ref{sec_dynamics} with convergence analysis in Section~\ref{sec_conv}. Simulations on economic dispatch and CPU scheduling appear in Section~\ref{sec_sim}. Section~\ref{sec_conclusion} concludes.

\paragraph{Notation}
$\mb{1}_n$ and $\mb{0}_n$ denote the $n$-vectors of ones and zeros. For $f\colon\mathbb{R}\to\mathbb{R}$, $\partial_x f := \tfrac{df}{dx}$. For $F\colon\mathbb{R}^n\to\mathbb{R}$, $\nabla F$ is its gradient. {We summarize the key notations in Table~\ref{table_notation}.}
\begin{table}[t]
	\centering
	\caption{Summary of the key notations in the paper.}\label{table_notation}
	\smallskip
	{
		\begin{tabular}{lc}
			\hline
			Symbol & Description\\
			\hline \hline 
			$f_i$,$F$
			& local and global objective \\
			\hline
			$x_i$
			& state variable of agent $i$  \\
			\hline
			$\mb{x}$
			& global state variable  \\
			\hline
			$n$
			& number of nodes/agents  \\
			\hline
			$b$
			& aggregate demand  \\
			\hline
			$\nu$,$\mu$
			& strong convexity and Lipschitz constant  \\
			\hline
			$\mb{x}^*$
			& global optimizer  \\
			\hline
			$\underline{m}_i$,$\overline{m}_i$
			& local box constraints  \\
			\hline
			$\sigma$
			& penalty weight term  \\
			\hline
			$f_i^\sigma$
			& penalty function \\
           \hline
           $W$,$L_\mc{G}$
           & adjacency and Laplacian matrix \\
            \hline
            $L_{\mathrm{sym}}$
            &  symmetric Laplacian matrix \\
            \hline
           $\lambda_2$,$\lambda_n$
           & smallest and largest nonzero eigenvalue of $L_{\mathrm{sym}}$\\
			\hline
			$\mc{S}_b$
			& feasible set\\
			\hline
			$k$
			& iteration time index\\
			\hline
			$\tau$
			& step size\\
			\hline
			$\pmb \phi$
			& $\mbox{sgn}(\nabla F)$\\
			\hline
			$\eta(\tau)$
			& descent threshold\\
			\hline
			$\sigma_L$ &  spectral norm of $L_{\mc{G}}$ \\
			\hline
			$\overline{F}$ &  optimality residual \\
			\hline
			$\widetilde{\mb{x}}$&  residual error \\
			\hline
			$\xi$&  gradient error \\
			\hline \hline
		\end{tabular}
	}
\end{table}

%%%%%%%%%%%%%%%%%%%%%%%%%%%%%%%%%%%%%%%%%%

\section{Problem Formulation} \label{sec_def}

Consider a network of $n$ agents indexed by $\mathcal{V} = \{1,\dots,n\}$. Each agent $i$ holds a local cost function $f_i\colon\mathbb{R}\to\mathbb{R}$ and controls a scalar decision variable $x_i\in\mathbb{R}$. We study the equality-constrained resource allocation problem
\begin{align} \label{eq_dra}
	\mathcal{P}_1:\quad \min_{\mb{x}\in\mathbb{R}^n} \; F(\mb{x}) := \sum_{i=1}^{n} f_i(x_i), \qquad \text{s.t.} \quad \sum_{i=1}^{n} x_i = b,
\end{align}
where $\mb{x} = (x_1,\dots,x_n)^\top$ and $b\in\mathbb{R}$ is the aggregate demand. {In this work, as in many resource-allocation and scheduling problems, we consider scalar resource variable to each agent. Two examples of the generated power of generator in economic dispatch and the CPU allocation of server in the scheduling are provided for the simulations.}

\begin{ass}[Cost regularity] \label{ass_strong}
	For every $i\in\mathcal{V}$, the local cost $f_i$ is twice continuously differentiable and $\nu$-strongly convex with $\mu$-Lipschitz continuous gradient, i.e., there exist constants $0 < \nu \leq \mu < \infty$ such that
	\begin{align}\label{eq_strong_smooth}
		\nu \;\leq\; \frac{d^2 f_i}{d x_i^2}(x_i) \;\leq\; \mu, \qquad \forall\, x_i\in\mathbb{R},\;\forall\, i\in\mathcal{V}.
	\end{align}
\end{ass}

Under Assumption~\ref{ass_strong}, problem $\mathcal{P}_1$ is strongly convex over the affine feasible set and, therefore, admits a unique minimizer $\mb{x}^*$ (cf.\ Lemmas~\ref{lem_optimal_solution}--\ref{lem_unique_feasible} below).

\paragraph{Box constraints via penalty relaxation}
In several applications (e.g., generator capacity limits in economic dispatch), the decision variables are additionally subject to local box constraints
\begin{align}\label{eq_box}
	\underline{m}_i \;\leq\; x_i \;\leq\; \overline{m}_i, \qquad i\in\mathcal{V}.
\end{align}
{Rather than enforcing~\eqref{eq_box} explicitly, we incorporate the box constraints through a quadratic hinge penalty \cite{bertsekas2003convex,nesterov1998introductory}. Define \begin{align}\label{eq_penalty} p_i(x_i) := [\underline{m}_i - x_i]_+^2 + [x_i - \overline{m}_i]_+^2, \qquad [z]_+:=\max\{z,0\}, 
\end{align} and set $f_i^\sigma := f_i + \sigma p_i$. 
\begin{rem}\label{rem_penalty_assumption} The derivative $p_i'(x) = -2[\underline{m}_i-x]_+ + 2[x-\overline{m}_i]_+$ consists of two terms with disjoint supports and is therefore globally $2$-Lipschitz. Hence $f_i^\sigma$ is convex and continuously differentiable but, for this quadratic hinge, \emph{not} twice continuously differentiable at the bounds. The constant $\nu$ remains a valid global strong-convexity constant for $f_i^\sigma$, and $(f_i^\sigma)'$ is $(\mu+2\sigma)$-Lipschitz, so a valid global condition-number bound is 
\begin{align}\label{eq_kappa_sigma} \kappa_\sigma \;\le\; \frac{\mu+2\sigma}{\nu}. 
\end{align}
\end{rem}
The above expression reveals that the condition number deteriorates as $\sigma$ increases.}
Some literature also suggest logarithmic-barrier function defined as $\tfrac{1}{\kappa}\log\!\bigl(1+\exp(\kappa u)\bigr)$~\cite{csl2021}, and $\theta$-logarithmic barriers~\cite{li2017distributed}.

\subsection{Network Model and Graph-Theoretic Preliminaries} \label{sec_graph}

The communication network at discrete time $k\geq 0$ is modeled as a directed graph (digraph) $\mc{G}(k)=(\mc{V},\mc{E}(k))$ with a fixed node set $\mc{V}=\{1,\dots,n\}$ and a time-varying edge set $\mc{E}(k)\subseteq \mc{V}\times\mc{V}$. A directed edge $(i,j)\in\mc{E}(k)$ indicates that node~$j$ can receive information from node~$i$ at time~$k$, and carries a positive weight $W_{ji}(k)>0$; we set $W_{ji}(k)=0$ whenever $(i,j)\notin\mc{E}(k)$. The matrix $W(k)=[W_{ij}(k)]\in\mathbb{R}_{\geq 0}^{n\times n}$ is the (weighted) adjacency matrix of $\mc{G}(k)$. Self-loops are excluded, i.e., $W_{ii}(k)=0$ for all $i$.

For each node $i$, denote by $\mc{N}_i^+(k)=\{j\in\mc{V}:(j,i)\in\mc{E}(k)\}$ its set of \emph{in-neighbours} (nodes from which $i$ receives) and by $\mc{N}_i^-(k)=\{j\in\mc{V}:(i,j)\in\mc{E}(k)\}$ its set of \emph{out-neighbours} (nodes to which $i$ sends).

\begin{defn}[Weight-balanced digraph]\label{def_wb}
	The digraph $\mc{G}(k)$ is \emph{weight-balanced} (WB) (or referred to just as balanced) if every node has equal weighted in- and out-degree, i.e., 
	\begin{align}\label{eq_wb}
		\sum_{j\in\mc{N}_i^+(k)} W_{ij}(k) \;=\; \sum_{j\in\mc{N}_i^-(k)} W_{ij}(k), \qquad \forall\, i\in\mc{V},
	\end{align}
	or, equivalently, $W(k)^\top\mb{1}_n = W(k)\mb{1}_n$.
\end{defn}

\begin{defn}[Strong connectivity]\label{def_sc}
	A digraph $\mc{G}$ is \emph{strongly connected} (SC) if, for every ordered pair $(i,j)\in\mc{V}\times\mc{V}$ with $i\neq j$, there exists a directed path from $i$ to $j$.
\end{defn}

\begin{ass}[Network connectivity] \label{ass_G}
	The following hold for all $k\geq 0$:
	\begin{enumerate}
		\item[(i)] $\mc{G}(k)$ is weight-balanced.
		\item[(ii)] There exists a finite integer $B\geq 1$ such that the \emph{union graph} $\mc{G}_B(k):=\bigl(\mc{V},\,\bigcup_{\ell=k}^{k+B-1}\mc{E}(\ell)\bigr)$ is strongly connected for every $k\geq 0$ (i.e., the network is \emph{uniformly} or \emph{$B$-connected}).
	\end{enumerate}
\end{ass}

\begin{rem}\label{rem_wb_vs_stochastic}
	Assumption~\ref{ass_G}(i) requires only that $W(k)^\top\mb{1}_n = W(k)\mb{1}_n$, which is strictly weaker than the doubly-stochastic condition $\mb{1}_n^\top W(k) = \mb{1}_n^\top,\; W(k)\mb{1}_n = \mb{1}_n$ assumed in~\cite{rikos2021optimal,falsone2020tracking,doan2017scl}. In particular, weight balance is preserved under arbitrary positive rescaling of edge weights, whereas doubly-stochastic weights must be explicitly redesigned whenever the topology changes -- a significant burden in time-varying or unreliable networks.
\end{rem}

The \emph{Laplacian} of $\mc{G}(k)$ is given by
\begin{align}\label{eq_laplacian_def}
	L_{\mc{G}}(k) \;:=\; D(k) - W(k),
\end{align}
where $D(k)=\operatorname{diag}\!\bigl(\sum_{j\in\mc{N}_i^-(k)}W_{ij}(k)\bigr)_{i=1}^{n}=\operatorname{diag}\!\bigl(\sum_{j\in\mc{N}_i^+(k)}W_{ij}(k)\bigr)_{i=1}^{n}$,   under the WB condition. We further define the \emph{symmetric part} $L_{\mathrm{sym}}(k):=\tfrac{1}{2}\bigl(L_{\mc{G}}(k)+L_{\mc{G}}(k)^\top\bigr)$. Lastly, for any $\mb{v}\in\mathbb{R}^n$, its \emph{mean-centered} version $\overline{\mb{v}}:=\mb{v}-\tfrac{\mb{1}_n^\top\mb{v}}{n}\mb{1}_n$.

\begin{lem}[Laplacian spectral properties {\cite{my_cpu}}] \label{lem_L}
	Let Assumption~\ref{ass_G} hold and let $\mb{y},\mb{z}\in\mathbb{R}^n$ satisfy $y_i z_i > 0$ for all $i\in\mc{V}$. Then, the following holds:
	\begin{enumerate}
		\item $L_{\mathrm{sym}}(k)$ is positive semi-definite with $L_{\mathrm{sym}}(k)\mb{1}_n = \mb{0}_n$, and the WB condition gives $L_{\mc{G}}(k)\mb{1}_n = L_{\mc{G}}(k)^\top\mb{1}_n = \mb{0}_n$.

		\item (Mean-centering invariance)
		\begin{align}\label{eq_Lmean}
			\mb{z}^\top L_{\mc{G}}(k)\,\mb{y} \;=\; \overline{\mb{z}}^\top L_{\mc{G}}(k)\,\overline{\mb{y}}.
		\end{align}

		\item (Spectral bounds) Let $\lambda_2(k)$ and $\lambda_n(k)$ denote the smallest and largest nonzero eigenvalues of $L_{\mathrm{sym}}(k)$, respectively (the Fiedler value~\cite{wu2005algebraic} and its upper counterpart). Then, we have
		\begin{align}\label{eq_spectral_bounds}
			\lambda_2(k)\;\overline{\mb{y}}^\top\overline{\mb{z}} \;\leq\; \overline{\mb{z}}^\top L_{\mc{G}}(k)\,\overline{\mb{y}} \;\leq\; \lambda_n(k)\;\overline{\mb{y}}^\top\overline{\mb{z}}.
		\end{align}
	\end{enumerate}
	All three statements remain valid with $L_{\mc{G}}(k)$ replaced by $L_{\mc{G}_B}(k)$ (the Laplacian of the union graph) and the corresponding eigenvalues $\lambda_{2,B}(k)$, $\lambda_{n,B}(k)$.
\end{lem}

\subsection{Convex Analysis Preliminaries} \label{sec_convex_prelim}

We collect several standard and known results on strongly convex resource allocation that underpin the convergence analysis in Section~\ref{sec_conv}. Throughout, we denote the \emph{feasible set} by
\begin{align}\label{eq_feasible_set}
	\mc{S}_b \;:=\; \bigl\{\mb{x}\in\mathbb{R}^n \;\big|\; \textstyle\sum_{i=1}^n x_i = b\bigr\},
\end{align}
the \emph{optimality residual} by $\overline{F}(\mb{x}):= F(\mb{x}) - F(\mb{x}^*)$, and the \emph{error vector} by $\widetilde{\mb{x}}:=\mb{x}-\mb{x}^*$.

\begin{lem}[KKT optimality {\cite{boyd2006optimal,Boyd-CVXBook}}] \label{lem_optimal_solution}
	Under Assumption~\ref{ass_strong}, problem $\mc{P}_1$ has a unique minimizer $\mb{x}^*\in\mc{S}_b$ characterized by equal marginal costs:
	\begin{align}\label{eq_equal_marginal}
		\partial_{x_i}f_i(x_i^*) \;=\; \partial_{x_j}f_j(x_j^*), \qquad \forall\, i,j\in\mc{V},
	\end{align}
	or, equivalently, $\nabla F(\mb{x}^*)\in\mathrm{span}(\mb{1}_n)$.
\end{lem}

\begin{lem}[Uniqueness on the feasible set {\cite{ccta22}}] \label{lem_unique_feasible}
	Under Assumption~\ref{ass_strong}, for every $b\in\mathbb{R}$ the point $\mb{x}^*$ furnished by Lemma~\ref{lem_optimal_solution} is the \emph{only} element of $\mc{S}_b$ satisfying $\nabla F(\mb{x}^*)\in\mathrm{span}(\mb{1}_n)$.
\end{lem}

\begin{lem}[Descent inequality {\cite{boyd2006optimal,Boyd-CVXBook}}] \label{lem_taylor}
	Let Assumption~\ref{ass_strong} hold. For any $\mb{x}_1,\mb{x}_2\in\mathbb{R}^n$ with $\delta\mb{x}:=\mb{x}_1-\mb{x}_2$,
	\begin{align}\label{eq_descent}
		F(\mb{x}_1) \;\geq\; F(\mb{x}_2) \;+\; \nabla F(\mb{x}_2)^\top\delta\mb{x} \;+\; \frac{\nu}{2}\,\|\delta\mb{x}\|_2^2.
	\end{align}
	Similarly, the $\mu$-smoothness upper bound gives
	\begin{align}\label{eq_ascent}
		F(\mb{x}_1) \;\leq\; F(\mb{x}_2) \;+\; \nabla F(\mb{x}_2)^\top\delta\mb{x} \;+\; \frac{\mu}{2}\,\|\delta\mb{x}\|_2^2.
	\end{align}
\end{lem}

\begin{rem}\label{rem_lemma4_direction}
	The original formulation states the bound~\eqref{eq_ascent} with constant $\mu$ (Lipschitz smoothness), which provides an \emph{upper} bound on the function change. This is the inequality used in the proof of Theorem~\ref{thm_converg} to establish sufficient descent. The complementary lower bound~\eqref{eq_descent} with constant $\nu$ (strong convexity) is used to relate the optimality residual to the error norm in Lemma~\ref{lem_cortes} below.
\end{rem}

\begin{lem}[Gradient and error bounds {\cite{cherukuri2015tcns,my_cpu}}] \label{lem_cortes}
	Let Assumption~\ref{ass_strong} hold and define the \emph{centered gradient}
	\begin{align}\label{eq_xi_def}
		\xi(\mb{x}) \;:=\; \nabla F(\mb{x}) \;-\; \frac{1}{n}\bigl(\textstyle\sum_{i=1}^n \partial_{x_i}f_i(x_i)\bigr)\mb{1}_n.
	\end{align}
	Then, for every $\mb{x}\in\mc{S}_b$, we obtain the following:
	\begin{enumerate}
		\item[(i)] \emph{(Residual--gradient equivalence)}
		\begin{align}\label{eq_residual_gradient}
			\frac{1}{2\mu}\,\|\xi(\mb{x})\|_2^2 \;\leq\; \overline{F}(\mb{x}) \;\leq\; \frac{1}{2\nu}\,\|\xi(\mb{x})\|_2^2.
		\end{align}

		\item[(ii)] \emph{(Residual--error equivalence)}
		\begin{align}\label{eq_residual_error}
			\frac{\nu}{2}\,\|\widetilde{\mb{x}}\|_2^2 \;\leq\; \overline{F}(\mb{x}) \;\leq\; \frac{\mu}{2}\,\|\widetilde{\mb{x}}\|_2^2.
		\end{align}

		\item[(iii)] \emph{(Gradient--error equivalence)}
		\begin{align}\label{eq_gradient_error}
			\frac{\|\xi(\mb{x})\|_2}{\mu} \;\leq\; \|\widetilde{\mb{x}}\|_2 \;\leq\; \frac{\|\xi(\mb{x})\|_2}{\nu}.
		\end{align}
	\end{enumerate}
\end{lem}

\section{The Single-Bit Allocation Algorithm} \label{sec_dynamics}

The key idea is to replace the full-precision gradient exchange in standard LG dynamics~\cite{cherukuri2015tcns,boyd2006optimal,gharesifard2013distributed} with a single-bit transmission: each agent broadcasts only $\mbox{sgn}(\partial_{x_i}f_i)$ to its out-neighbours. Concretely, under the standard protocol every agent sends a 64-bit floating-point value per link per iteration, yielding a network-wide payload of $64\,|\mc{E}(k)|$ bits per round. The proposed scheme reduces this to $|\mc{E}(k)|$ bits. Thus, a $64\times$ reduction that grows linearly with network size and link density; for instance, a network of $n{=}1000$ agents with average out-degree~$10$ saves approximately $640\,$kbits per iteration.

We propose the following discrete-time dynamics to solve $\mc{P}_1$: each agent~$i$ updates its state as
\begin{equation} \label{eq_sol}
	x_i (k+1) = x_i(k) - \tau\sum_{j \in \mc{N}^-_i} W_{ij} \Big(\mbox{sgn}(\partial_{x_i}f_i(k)) - \mbox{sgn}(\partial_{x_j}f_j(k))\Big),
\end{equation}
where $\partial_{x_i}f_i(k)$ denotes the derivative of $f_i$ evaluated at $x_i(k)$ and $\tau>0$ is a fixed step-size. In compact form, writing $\pmb \phi(k):=\mbox{sgn}(\nabla F(k))$ element-wise, we can rewrite~\eqref{eq_sol} in the matricial form as
\begin{align} \label{eq_sol_L}
	\mb{x}(k+1) = \mb{x}(k) -
	\tau L_{\mc{G}}(k)\, \pmb \phi(k),
\end{align}
and when the graph is static we suppress the time argument and write $L_{\mc{G}}$.

\paragraph{Relation to existing nonlinear protocols}
Signum-based distributed optimization and allocation protocols have appeared in~\cite{parsegov2013fixed,spl,chen2022distributed}, using the power mapping $\mbox{sgn}^v(\mb{x})=\mb{x}\|\mb{x}\|^{v-1}$ ($v>0$) to achieve finite- or fixed-time convergence. Similarly, sign-preserving consensus dynamics resilient to impulsive noise are studied in~\cite{wei2017consensus,stankovic2020nonlinear}. The distinguishing feature of~\eqref{eq_sol} is that only a \emph{single bit} (i.e., the sign $\{+1,-1\}$) is transmitted per link, whereas the schemes in~\cite{parsegov2013fixed,spl,chen2022distributed,gharesifard2013distributed,cherukuri2015tcns,shames2011accelerated} require multi-bit (or full-precision) exchange.

\paragraph{Structural properties}
Two properties of~\eqref{eq_sol} deserve emphasis before the formal analysis in Section~\ref{sec_conv}. First, because the update has a Laplacian structure ($\mb{1}_n^\top L_{\mc{G}}=\mb{0}_n^\top$ under weight balance), the sum $\sum_i x_i(k)$ is preserved at every iteration, guaranteeing \emph{anytime feasibility}, which is an improvement over primal-dual and ADMM-based methods~\cite{li2017distributed,dominguez2012decentralized,kar2012distributed,turan2020resilient,nesterov2018dual,feijer2010stability} that achieve feasibility only asymptotically. Second, only the weight-balance condition (Assumption~\ref{ass_G}) is needed, not doubly-stochastic weights; this avoids costly weight re-design after link failures~\cite{ejc}.

The complete procedure is listed in Algorithm~\ref{alg_1}.

\begin{algorithm}
	\textbf{Input:}  $W$, $\mc{N}^-_i$, $\mc{N}^+_i$, $\tau$, $f_i(\cdot)$\;
	\textbf{Initialization:}  node $i$ randomly sets $\underline{m}_i \leq x_i(0) \leq \overline{m}_i$ satisfying $\sum_i x_i(0)=b$ and  $\mbox{sgn}(\partial_{x_i}f_i(0)) \neq \mbox{sgn}(\partial_{x_j}f_j(0))$ for at least one $j \in \mc{N}_i^+$ or $j \in \mc{N}_i^-$ {(this is a one-time procedure)}. Set $k=0$ \; 
	\While{termination criteria NOT met}{
		Each node $i$ receives a single-bit packet from $j \in \mc{N}^-_i$ (either $+1$ or $-1$) and computes Eq.~\eqref{eq_sol}\;
		\If{no packet from $j \in \mc{N}^-_i$}{sets $W_{ij}=0$ in Eq.~\eqref{eq_sol}}
		\If{$\mbox{sgn}(\partial_{x_i}f_i(k)) = +1$}{sends the bit $+1$ to nodes $j \in \mc{N}^+_i$}
		\Else{\If{$\mbox{sgn}(\partial_{x_i}f_i(k)) = -1$}{sends the bit $-1$ to nodes $j \in \mc{N}^+_i$}}
		Sets $k \leftarrow k+1$\;
	}
	\textbf{Return} Final state $\mb{x}(k)$ and cost $F(\mb{x}(k))$	
	\caption{Single-Bit Resource Allocation}
	\label{alg_1}
\end{algorithm} \normalsize

\paragraph{Initialization}
Feasible initialization requires $\sum_i x_i(0)=b$ together with $\mbox{sgn}(\partial_{x_i}f_i(0)) \neq \mbox{sgn}(\partial_{x_j}f_j(0))$ for at least one neighbour pair, ensuring $x_i(1)\neq x_i(0)$ under~\eqref{eq_sol}. The simplest choice is $x_i(0)=b/n$ for all~$i$; for heterogeneous box constraints a spanning-tree message-passing routine~\cite[Algorithm~2]{cherukuri2015tcns} can be used. {We note that this initialization phase itself requires multi-bit communication, but it is executed \emph{once} before the iterative single-bit loop begins and its cost is paid off over the (potentially very large) number of subsequent iterations.}

\begin{rem}\label{rem_improvements}
	Relative to the existing literature, Algorithm~\ref{alg_1} simultaneously provides three relaxations: (i)~all-time connectivity is weakened to $B$-connecti\-vity (Assumption~\ref{ass_G}(ii)), accommodating intermittent link failures; (ii)~dou\-bly-stochastic weights are replaced by the weaker weight-balance condition, eliminating the need for compensation algorithms upon topology changes~\cite{vaidya2012robust}; and (iii)~per-link communication is reduced from multi-bit to a single bit. These improvements hold concurrently, whereas existing methods~\cite{dominguez2012decentralized,li2017distributed,doan2017scl,boyd2006optimal,zhou2024fixed,kar_edp} require stronger assumptions in at least one of these dimensions.
\end{rem}
{
\begin{rem}
	In the proposed resource allocation technique in Algorithm~\ref{alg_1}, each active link transmits one sign bit per iteration; a packet loss corresponds to an inactive link (i.e., $W_{ij}(k)=0$). No zero-gradient value is inferred from a missing packet; and the switching-topology results cover this packet-loss interpretation.
\end{rem}}
\section{Convergence Analysis} \label{sec_conv}

This section establishes the convergence properties of the single-bit dynamics~\eqref{eq_sol}. The argument proceeds in four stages. First, we show that the feasible set $\mc{S}_b$ is positively invariant, so the equality constraint is preserved at every iterate (Lemma~\ref{lem_feasible_intime}). Second, we characterize the invariant set $\mc{I}$ of the dynamics and prove that it consists precisely of those feasible points at which all gradient signs agree (Lemma~\ref{lem_tree}). Third, using $\overline{F}(\mb{x}) = F(\mb{x}) - F(\mb{x}^*)$ as a Lyapunov function, we show that $\overline{F}$ decreases monotonically whenever the gradient norm exceeds a $\tau$-dependent threshold, establishing finite-time entry into a neighborhood of~$\mc{I}$ whose size scales as $O(\tau)$ (Theorem~\ref{thm_converg}). Fourth, we quantify the steady-state optimality gap and error bounds within this neighborhood.

\begin{lem}[All-time feasibility] \label{lem_feasible_intime}
	Let Assumption~\ref{ass_G} hold and $\mb{x}(0) \in \mc{S}_b$. Then $\mb{x}(k) \in \mc{S}_b$ for every $k \geq 0$ under the dynamics~\eqref{eq_sol}. In particular, the set $\mc{S}_b$ is positively invariant.
\end{lem}
\begin{proof}
	From the compact form~\eqref{eq_sol_L},
	\begin{align}\label{eq_feas_proof}
		\mb{1}_n^\top \mb{x}(k+1)
		= \mb{1}_n^\top \mb{x}(k) - \tau\, \mb{1}_n^\top L_{\mc{G}}(k)\, \mbox{sgn}\!\bigl(\nabla F(k)\bigr).
	\end{align}
	Under the weight-balance condition (Assumption~\ref{ass_G}(i)), Lemma~\ref{lem_L} gives $\mb{1}_n^\top L_{\mc{G}}(k) = \mb{0}_n^\top$, so the second term vanishes. Induction on $k$ yields $\mb{1}_n^\top \mb{x}(k) = \mb{1}_n^\top \mb{x}(0) = b$ for all $k \geq 0$.
\end{proof}

An immediate and practically important consequence is \emph{anytime feasibility}: the equality constraint $\sum_i x_i(k) = b$ holds at every iteration, not merely in the limit. This is a structural improvement over ADMM-based and primal--dual methods~\cite{dominguez2012decentralized,kar2012distributed,li2017distributed}, which recover feasibility only asymptotically and therefore cannot guarantee constraint satisfaction upon early termination.

Next, we characterize the invariant set of~\eqref{eq_sol}. A point $\mb{x} \in \mc{S}_b$ is called \emph{invariant} (or stationary) if $\mb{x}(k) = \mb{x}$ implies $\mb{x}(k+1) = \mb{x}$ for all $k \geq 0$.

\begin{lem}[Invariant set]\label{lem_tree}
	Let Assumptions~\ref{ass_strong}--\ref{ass_G} hold and let $\mb{x}(0)\in\mc{S}_b$
	satisfy the initialization conditions of Algorithm~\ref{alg_1}. Then the invariant
	set of~\eqref{eq_sol} is
	\begin{align}\label{eq_inv_set}
		\mc{I}\;=\;\bigl\{\mb{x}\in\mc{S}_b \;\big|\; \mbox{sgn}\!\bigl(\nabla F(\mb{x})\bigr)\in \mathrm{span}(\mb{1}_n)\bigr\},
	\end{align}
	where $\mbox{sgn}$ is applied element-wise. In particular, the optimizer
	$\mb{x}^*$ satisfies $\mb{x}^*\in\mc{I}$.
\end{lem}
\begin{proof}
	Let $\pmb{\psi}(\mb{x}) := \mbox{sgn}(\nabla F(\mb{x}))\in\{-1,0,+1\}^n$.

	\smallskip\noindent\underline{Sufficiency.}
	Take any $\mb{x}\in\mc{S}_b$ such that $\pmb{\psi}(\mb{x})\in\mathrm{span}(\mb{1}_n)$.
	Then $\pmb{\psi}(\mb{x})=c\,\mb{1}_n$ for some $c\in\{-1,0,+1\}$. Using the Laplacian
	form~\eqref{eq_sol_L} and $L_{\mc{G}}(k)\mb{1}_n=\mb{0}_n$
	(Lemma~\ref{lem_L}), we obtain
	$\mb{x}(k+1)=\mb{x}(k)-\tau L_{\mc{G}}(k)\,c\mb{1}_n=\mb{x}(k)$ for all $k$,
	so $\mb{x}$ is invariant.

	\smallskip\noindent\underline{Necessity.}
	Assume, for contradiction, that $\mb{x}\in\mc{S}_b$ is invariant under~\eqref{eq_sol}
	but $\pmb{\psi}:=\pmb{\psi}(\mb{x})\notin\mathrm{span}(\mb{1}_n)$. Invariance means
	$\mb{x}(k)=\mb{x}$ for all $k$, hence $\pmb{\psi}(\mb{x}(k))=\pmb{\psi}$ for all $k$.
	Define the minimum sign value and its index set
	\begin{align*}
		\psi_{\min}:=\min_{i\in\mc{V}}\psi_i,\qquad \mc{M}:=\{i\in\mc{V}:\psi_i=\psi_{\min}\}.
	\end{align*}
	Because $\pmb{\psi}\notin\mathrm{span}(\mb{1}_n)$, the set $\mc{M}$ is a nonempty proper subset of $\mc{V}$.

	Fix any time $k\ge 0$ and consider the union graph over a $B$-window,
	$\mc{G}_B(k)=\bigl(\mc{V},\cup_{\ell=k}^{k+B-1}\mc{E}(\ell)\bigr)$, which is strongly
	connected by Assumption~\ref{ass_G}(ii). Pick any node $r\in\mc{V}\setminus\mc{M}$ and any
	node $q\in\mc{M}$. Since $\mc{G}_B(k)$ is strongly connected, there exists a directed path
	from $r$ to $q$ in $\mc{G}_B(k)$. Along that path, let $(p,\bar q)$ be the \emph{first} edge
	that enters $\mc{M}$, i.e., $p\notin\mc{M}$ and $\bar q\in\mc{M}$. Then
	$(p,\bar q)\in \bigcup_{\ell=k}^{k+B-1}\mc{E}(\ell)$,
	so there exists $k^*\in\{k,\dots,k+B-1\}$ such that $(p,\bar q)\in\mc{E}(k^*)$.

	At time $k^*$, node $\bar q$ updates according to~\eqref{eq_sol}, i.e., 
	\begin{align*}
		x_{\bar q}(k^*+1)-x_{\bar q}(k^*)
		= -\tau \sum_{j\in\mc{N}_{\bar q}^-(k^*)} W_{\bar q j}(k^*)\,\bigl(\psi_{\bar q}-\psi_j\bigr).
	\end{align*}
	For every in-neighbor $j\in\mc{N}_{\bar q}^-(k^*)$ we have $\psi_j\ge \psi_{\min}=\psi_{\bar q}$,
	hence $(\psi_{\bar q}-\psi_j)\le 0$ and therefore each summand satisfies
	$-\tau\,W_{\bar q j}(k^*)\,(\psi_{\bar q}-\psi_j) \ge 0$.
	Moreover, since $p\notin\mc{M}$ we have $\psi_p>\psi_{\bar q}$, so $(\psi_{\bar q}-\psi_p)<0$
	and with $W_{\bar q p}(k^*)>0$ the corresponding summand is \emph{strictly} positive.
	Because all summands are non-negative and at least one is strictly positive,
	\begin{align*}
		x_{\bar q}(k^*+1)-x_{\bar q}(k^*)\ >\ 0,
	\end{align*}
	which contradicts invariance ($x_{\bar q}(k^*+1)=x_{\bar q}(k^*)$). Therefore,
	no invariant $\mb{x}\in\mc{S}_b$ can satisfy $\pmb{\psi}(\mb{x})\notin\mathrm{span}(\mb{1}_n)$.

	Lastly, notice that by Lemma~\ref{lem_optimal_solution}, $\nabla F(\mb{x}^*)\in\mathrm{span}(\mb{1}_n)$,
	hence $\pmb{\psi}(\mb{x}^*)\in\mathrm{span}(\mb{1}_n)$ and, thus, $\mb{x}^*\in\mc{I}$.
\end{proof}

\begin{rem}[Role of uniform connectivity]\label{rem_invariant_Bconnected}
	The necessity argument crucially relies on the $B$-connectivity of the union graph. At any single time instant $k$, the instantaneous graph $\mc{G}(k)$ may be disconnected, and the kernel of $L_{\mc{G}}(k)$ may contain vectors outside $\mathrm{span}(\mb{1}_n)$; in such cases $L_{\mc{G}}(k)\,\pmb{\psi} = \mb{0}_n$ does not imply $\pmb{\psi} \in \mathrm{span}(\mb{1}_n)$. The $B$-window argument in Step~2 circumvents this: it shows that within \emph{every} window of $B$ consecutive steps, the union graph is strongly connected and exposes at least one disagreement edge, which produces a non-zero displacement and rules out invariance. In other words, the invariant-set characterization~\eqref{eq_inv_set} holds for the dynamics~\eqref{eq_sol} driven by the time-varying sequence $\{\mc{G}(k)\}_{k\geq 0}$, provided the union over every $B$-window is strongly connected.
\end{rem}

\begin{rem}[Generality of the sign nonlinearity]\label{rem_general_nonlinearity}
	{Both the feasibility result (Lemma~\ref{lem_feasible_intime}) and the invariant-set characterization (Lemma~\ref{lem_tree}) hold if $\mbox{sgn}(\cdot)$ is replaced by \emph{any} odd, sign-preserving mapping $\varphi\colon\mathbb{R}\to\mathbb{R}$ satisfying $u\,\varphi(u) > 0$ for $u \neq 0$ and $\varphi(0) = 0$. The Laplacian annihilation property $\mb{1}_n^\top L_{\mc{G}} = \mb{0}_n^\top$ depends only on the weight-balance structure, not on the specific form of $\varphi$, and the disagreement argument in Step~2 of Lemma~\ref{lem_tree} requires only that $\varphi(a) \neq \varphi(b)$ whenever $\mbox{sgn}(a) \neq \mbox{sgn}(b)$. The signum function is the extreme one-bit instance of this class.}
\end{rem}

Having established feasibility and the invariant set, we proceed to the Lyapunov-based analysis of descent and neighborhood convergence. We first record a minor extension of the spectral bound in Lemma~\ref{lem_L} that accommodates zero-gradient components, then state and prove the main convergence result.

\begin{lem}[Extension of Lemma~\ref{lem_L}(3) to non-strict sign alignment]\label{lem_L_extended}
	The spectral bounds in Lemma~\ref{lem_L}(3) continue to hold if the hypothesis is relaxed to $y_iz_i \geq 0$ for all $i\in\mc{V}$ (with $\overline{\mb{y}} \neq \mb{0}_n$ and $\overline{\mb{z}} \neq \mb{0}_n$).
\end{lem}
\begin{proof}
	For any $\delta > 0$, define $\mb{y}^\delta := \mb{y} + \delta\,\mb{z}$. Then $y_i^\delta z_i = y_iz_i + \delta z_i^2 > 0$ for every $i$ with $z_i \neq 0$, and $y_i^\delta z_i = 0$ only if $z_i = 0$ (in which case both sides of the spectral bound vanish for that component). Since $\overline{\mb{z}} \neq \mb{0}_n$, there exists $i$ with $z_i \neq 0$, so the strict hypothesis of Lemma~\ref{lem_L}(3) holds for $(\mb{y}^\delta, \mb{z})$ restricted to the nonzero-$z$ components. Applying Lemma~\ref{lem_L}(3) and taking $\delta \to 0^+$, the bound follows by continuity of both sides in $\mb{y}$.
\end{proof}

\begin{theorem}[Convergence to a $\tau$-dependent neighborhood under fixed step-size]\label{thm_converg}
Let Assumptions~\ref{ass_strong}--\ref{ass_G} hold with a \emph{static}, strongly connected,
weight-balanced graph~$\mc{G}$ and Laplacian $L_{\mc{G}}$. Let $\mb{x}(0)\in\mc{S}_b$
satisfy the initialization conditions of Algorithm~\ref{alg_1}, and define
\[
\pmb{\phi}(k):=\mbox{sgn}\!\bigl(\nabla F(\mb{x}(k))\bigr)\in\{-1,0,+1\}^n \qquad \text{(element-wise)}.
\]
Let $L_{\mathrm{sym}}:=\tfrac{1}{2}(L_{\mc{G}}+L_{\mc{G}}^\top)$, let $\lambda_2>0$ be the smallest nonzero eigenvalue of $L_{\mathrm{sym}}$,
and let $\sigma_L := \|L_{\mc{G}}\|_2$ denote the spectral norm (largest singular value) of $L_{\mc{G}}$.
For any fixed $\tau>0$, define the \emph{descent threshold}
\begin{align}\label{eq_eta_tau}
\eta(\tau)\;:=\;\frac{\mu\,n^2\,\sigma_L^2}{2\,\lambda_2}\;\tau.
\end{align}
Then, the following hold:
\begin{enumerate}
\item[\emph{(i)}] \emph{(Monotone descent outside the threshold region)}
For every $k$ such that $\pmb{\phi}(k)\notin\mathrm{span}(\mb{1}_n)$ and
$\|\nabla F(\mb{x}(k))\|_1\ge \eta(\tau)$, the optimality residual
$\overline{F}(k):=F(\mb{x}(k))-F(\mb{x}^*)$ satisfies
\[
\overline{F}(k+1)\;\le\; \overline{F}(k).
\]
Moreover, for any $\varepsilon>0$, if $\|\nabla F(\mb{x}(k))\|_1\ge \eta(\tau)+\varepsilon$, then
\begin{align}\label{eq_uniform_drop}
\overline{F}(k+1)\;\le\; \overline{F}(k)\;-\;\frac{\lambda_2\,\tau}{n}\,\varepsilon.
\end{align}

\item[\emph{(ii)}] \emph{(Finite excursion bound)}
Fix any $\varepsilon>0$. Then either $\mb{x}(k)$ reaches the invariant set
$\mc{I}=\{\mb{x}\in\mc{S}_b:\pmb{\phi}(\mb{x})\in\mathrm{span}(\mb{1}_n)\}$
at some finite time, or else the \emph{total number} of iterations~$k$ at which
$\|\nabla F(\mb{x}(k))\|_1 \ge \eta(\tau)+\varepsilon$ is at most
\begin{align}\label{eq_excursion_bound}
K_\varepsilon \;:=\; \Bigl\lceil \frac{n\,\overline{F}(0)}{\lambda_2\,\tau\,\varepsilon}\Bigr\rceil.
\end{align}
In particular, the iterates enter the set $\bigl\{\mb{x}\in\mc{S}_b : \|\nabla F(\mb{x})\|_1 < \eta(\tau)+\varepsilon\bigr\}$ for the first time no later than iteration~$K_\varepsilon$.

\item[\emph{(iii)}] \emph{(Neighborhood bounds when inside the threshold region)}
At any iteration $k$ satisfying $\|\nabla F(\mb{x}(k))\|_1 \le \eta(\tau)+\varepsilon$, the centered gradient
$\xi(\mb{x}(k))$ from~\eqref{eq_xi_def} satisfies
\[
\|\xi(\mb{x}(k))\|_2 \;\le\;\Bigl(1+\tfrac{1}{\sqrt{n}}\Bigr)\bigl(\eta(\tau)+\varepsilon\bigr),
\]
and therefore, by Lemma~\ref{lem_cortes}(i)--(iii),
\begin{align}\label{eq_gap_bounds}
&\overline{F}(k)\;\le\;\frac{(1+1/\!\sqrt{n})^2}{2\nu}\,\bigl(\eta(\tau)+\varepsilon\bigr)^2, \notag
\\ &\qquad \text{ and } \qquad
\|\mb{x}(k)-\mb{x}^*\|_2 \;\le\;\frac{1+1/\!\sqrt{n}}{\nu}\,\bigl(\eta(\tau)+\varepsilon\bigr).
\end{align}
\end{enumerate}
\end{theorem}

\begin{proof}
Define the one-step displacement $\delta\mb{x}:=\mb{x}(k+1)-\mb{x}(k) = -\tau\, L_{\mc{G}}\,\pmb{\phi}(k)$ from~\eqref{eq_sol_L}.

\smallskip\noindent\underline{Step~1: Smoothness bound.}
By $\mu$-smoothness (Lemma~\ref{lem_taylor}, Eq.~\eqref{eq_ascent}),
\begin{align}\label{eq_smooth_bound}
\overline{F}(k+1)-\overline{F}(k)
\;\le\;
\nabla F(k)^\top \delta\mb{x} + \tfrac{\mu}{2}\|\delta\mb{x}\|_2^2
\;=\;
-\tau\,\nabla F(k)^\top L_{\mc{G}}\,\pmb{\phi}(k)
\;+\;\tfrac{\mu}{2}\,\tau^2\,\pmb{\phi}(k)^\top L_{\mc{G}}^\top L_{\mc{G}}\,\pmb{\phi}(k).
\end{align}

\smallskip\noindent\underline{Step~2: Lower bound on the descent term.}
Write $\bar{g}(k) := \tfrac{1}{n}\sum_{i=1}^n \partial_{x_i}f_i(k)$ for the scalar gradient mean and $\bar{\phi}(k) := \tfrac{1}{n}\sum_{i=1}^n \phi_i(k)$ for the scalar sign mean. Since the graph is weight-balanced, we have $L_{\mc{G}}\,\mb{1}_n = \mb{0}_n$ and $\mb{1}_n^\top L_{\mc{G}} = \mb{0}_n^\top$, which gives the mean-centering identity (Lemma~\ref{lem_L}(2)):
\begin{align}\label{eq_centering}
\nabla F(k)^\top L_{\mc{G}}\,\pmb{\phi}(k)
\;=\;
\overline{\nabla F(k)}^\top L_{\mc{G}}\,\overline{\pmb{\phi}}(k),
\end{align}
where $\overline{\mb{v}} := \mb{v} - \tfrac{1}{n}(\mb{1}_n^\top\mb{v})\mb{1}_n$ denotes the mean-centered version of any vector~$\mb{v}$.

We apply Lemma~\ref{lem_L}(3) with $\mb{y} = \pmb{\phi}(k)$ and $\mb{z} = \nabla F(k)$; the lemma mean-centers these vectors internally and yields the spectral bound $\overline{\nabla F(k)}^\top L_{\mc{G}}\,\overline{\pmb{\phi}}(k) \geq \lambda_2\,\overline{\pmb{\phi}}(k)^\top\overline{\nabla F(k)}$. The sign hypothesis $y_iz_i = \phi_i(k)\cdot\partial_{x_i}f_i(k) = |\partial_{x_i}f_i(k)| \geq 0$ holds for every~$i$, with strict inequality on all components where $\partial_{x_i}f_i(k) \neq 0$. On components where $\partial_{x_i}f_i(k) = 0$ we have $y_i = z_i = 0$, so the strict hypothesis $y_iz_i > 0$ of Lemma~\ref{lem_L}(3) may fail; Lemma~\ref{lem_L_extended} extends the bound to this non-strict case. We obtain
\begin{align}\label{eq_spectral_descent}
\overline{\nabla F(k)}^\top L_{\mc{G}}\,\overline{\pmb{\phi}}(k)
\;\geq\;
\lambda_2\;\overline{\pmb{\phi}}(k)^\top\overline{\nabla F(k)}.
\end{align}
Define the \emph{effective descent coefficient}
\begin{align}\label{eq_gamma_def}
\gamma(k)
\;:=\;
\overline{\pmb{\phi}}(k)^\top\overline{\nabla F(k)}
&\;=\;
\nabla F(k)^\top \pmb{\phi}(k) \;-\; n\,\bar{g}(k)\,\bar{\phi}(k)\\
&\;=\;
\|\nabla F(k)\|_1 \;-\; n\,\bar{g}(k)\,\bar{\phi}(k),
\end{align}
where the second equality expands the mean-centering and the third uses $\nabla F(k)^\top\pmb{\phi}(k) = \sum_i |\partial_{x_i}f_i(k)| = \|\nabla F(k)\|_1$.

We bound $\gamma(k)$ from below. Since $|\bar{g}(k)| = \tfrac{1}{n}|\sum_i\partial_{x_i}f_i(k)| \leq \tfrac{1}{n}\|\nabla F(k)\|_1$ and $|\bar{\phi}(k)|\leq 1$, it readily follows that
\begin{align}\label{eq_gamma_lower}
\gamma(k)
\;\geq\;
\|\nabla F(k)\|_1\bigl(1-|\bar{\phi}(k)|\bigr).
\end{align}
If $\pmb{\phi}(k)\in\{-1,0,+1\}^n$ and $\pmb{\phi}(k)\notin\mathrm{span}(\mb{1}_n)$, at least one component of $\pmb{\phi}(k)$ differs from the others. The maximum of $|\bar{\phi}(k)|$ subject to this constraint is achieved when $n-1$ entries are equal and one differs (e.g., $(+1,\dots,+1,0)$), giving $|\bar{\phi}(k)| \leq 1-\tfrac{1}{n}$. Therefore, we obtain
\begin{align}\label{eq_gamma_uniform}
\gamma(k) \;\geq\; \frac{1}{n}\,\|\nabla F(k)\|_1.
\end{align}
Combining~\eqref{eq_centering}, \eqref{eq_spectral_descent}, and~\eqref{eq_gamma_uniform}, we have
\begin{align}\label{eq_descent_final}
\nabla F(k)^\top L_{\mc{G}}\,\pmb{\phi}(k) \;\geq\; \frac{\lambda_2}{n}\,\|\nabla F(k)\|_1.
\end{align}

\smallskip\noindent\underline{Step~3: Upper bound on the quadratic term.}
Since $L_{\mc{G}}\,\mb{1}_n = \mb{0}_n$, we have $L_{\mc{G}}\,\pmb{\phi}(k) = L_{\mc{G}}\,\overline{\pmb{\phi}}(k)$ and $\|\overline{\pmb{\phi}}(k)\|_2^2 \leq \|\pmb{\phi}(k)\|_2^2 \leq n$. Therefore, we get
\begin{align}\label{eq_quad_upper}
\pmb{\phi}(k)^\top L_{\mc{G}}^\top L_{\mc{G}}\,\pmb{\phi}(k)
\;=\;
\|L_{\mc{G}}\,\overline{\pmb{\phi}}(k)\|_2^2
\;\leq\;
\sigma_L^2\,\|\overline{\pmb{\phi}}(k)\|_2^2
\;\leq\;
n\,\sigma_L^2.
\end{align}

\smallskip\noindent\underline{Step~4: Descent outside the threshold (proof of part~(i)).}
Substituting~\eqref{eq_descent_final} and~\eqref{eq_quad_upper} into~\eqref{eq_smooth_bound}, we have
\begin{align}\label{eq_combined_bound}
\overline{F}(k+1)-\overline{F}(k)
\;\leq\;
-\frac{\lambda_2\,\tau}{n}\,\|\nabla F(k)\|_1 \;+\; \frac{\mu}{2}\,n\,\sigma_L^2\,\tau^2.
\end{align}
The right-hand side is non-positive whenever $\|\nabla F(k)\|_1 \geq \eta(\tau)$ with $\eta(\tau)$ as in~\eqref{eq_eta_tau}, since
\[
\frac{\lambda_2\,\tau}{n}\,\eta(\tau)
\;=\;
\frac{\lambda_2\,\tau}{n}\cdot\frac{\mu\,n^2\,\sigma_L^2}{2\,\lambda_2}\,\tau
\;=\;
\frac{\mu}{2}\,n\,\sigma_L^2\,\tau^2.
\]
This proves $\overline{F}(k+1)\leq\overline{F}(k)$.

If instead $\|\nabla F(k)\|_1 \geq \eta(\tau)+\varepsilon$, then~\eqref{eq_combined_bound} gives
\begin{align*}
\overline{F}(k+1)-\overline{F}(k)
\;\leq\;
-\frac{\lambda_2\,\tau}{n}\bigl(\eta(\tau)+\varepsilon\bigr) + \frac{\mu}{2}\,n\,\sigma_L^2\,\tau^2
\;=\;
-\frac{\lambda_2\,\tau}{n}\,\varepsilon,
\end{align*}
establishing~\eqref{eq_uniform_drop}. This completes the proof of part~(i).

\smallskip\noindent\underline{Step~5: Finite excursion bound (proof of part~(ii)).}
We prove both the first-entry time and the total excursion count.

\emph{First-entry time.} Let $k^*$ denote the first iteration at which either $\mb{x}(k^*) \in \mc{I}$ or $\|\nabla F(\mb{x}(k^*))\|_1 < \eta(\tau)+\varepsilon$. At every iteration $k < k^*$, neither condition holds, so $\pmb{\phi}(k)\notin\mathrm{span}(\mb{1}_n)$ and $\|\nabla F(\mb{x}(k))\|_1 \geq \eta(\tau)+\varepsilon$. By part~(i), the drop bound~\eqref{eq_uniform_drop} applies at each such~$k$, and since all pre-entry iterations are consecutive descent steps, the bound telescopes:
\[
\overline{F}(k^*) \;\leq\; \overline{F}(0) - k^*\cdot\frac{\lambda_2\,\tau}{n}\,\varepsilon.
\]
Combined with $\overline{F}(k^*) \geq 0$, this gives $k^* \leq \frac{n\,\overline{F}(0)}{\lambda_2\,\tau\,\varepsilon}$.

\emph{Total excursion count.} Define $\mc{K}_\varepsilon := \{k \geq 0 : \|\nabla F(\mb{x}(k))\|_1 \geq \eta(\tau)+\varepsilon \text{ and } \mb{x}(k)\notin\mc{I}\}$. At every $k \in \mc{K}_\varepsilon$, the drop bound~\eqref{eq_uniform_drop} gives $\overline{F}(k) - \overline{F}(k+1) \geq \frac{\lambda_2\,\tau}{n}\varepsilon$. For any $T > 0$,
\[
\overline{F}(0) - \overline{F}(T)
\;\geq\;
\sum_{\substack{k \in \mc{K}_\varepsilon \\ k < T}} \bigl(\overline{F}(k) - \overline{F}(k+1)\bigr)
\;\geq\;
|\mc{K}_\varepsilon \cap \{0,\dots,T{-}1\}| \cdot \frac{\lambda_2\,\tau}{n}\,\varepsilon,
\]
where the first inequality holds because every term in the sum is a positive contribution to the total decrease of $\overline{F}$ from $0$ to $T$ (increases at non-excursion steps are absorbed into $\overline{F}(T) \geq 0$ on the left-hand side). Letting $T \to \infty$ and using $\overline{F}(T) \geq 0$ yields $|\mc{K}_\varepsilon| \leq \frac{n\,\overline{F}(0)}{\lambda_2\,\tau\,\varepsilon}$, giving~\eqref{eq_excursion_bound}. This proves part~(ii).

\smallskip\noindent\underline{Step~6: Neighborhood bounds (proof of part~(iii)).}
At any iteration with $$\|\nabla F(\mb{x}(k))\|_1 \leq \eta(\tau)+\varepsilon,$$ we can write $\xi(k) = \nabla F(k) - \bar{g}(k)\,\mb{1}_n$ with scalar mean $\bar{g}(k) = \tfrac{1}{n}\mb{1}_n^\top\nabla F(k)$. Thus, we obtain
\begin{align*}
\|\xi(k)\|_2
&\;\leq\;
\|\nabla F(k)\|_2 + \sqrt{n}\,|\bar{g}(k)|
\;\leq\;
\|\nabla F(k)\|_1 + \frac{1}{\sqrt{n}}\,\|\nabla F(k)\|_1\\
&\;=\;
\Bigl(1+\frac{1}{\sqrt{n}}\Bigr)\|\nabla F(k)\|_1
\;\leq\;
\Bigl(1+\frac{1}{\sqrt{n}}\Bigr)\bigl(\eta(\tau)+\varepsilon\bigr),
\end{align*}
where we used $\|\mb{v}\|_2 \leq \|\mb{v}\|_1$, and $|\bar{g}(k)| \leq \tfrac{1}{n}\|\nabla F(k)\|_1$ with $\sqrt{n}\cdot\tfrac{1}{n} = \tfrac{1}{\sqrt{n}}$. Applying Lemma~\ref{lem_cortes}(i) and~(iii) yields~\eqref{eq_gap_bounds}, completing the proof of part~(iii).
\end{proof}

{
Simply speaking, the following overview of Theorem~\ref{thm_converg} summarizes the main convergence results:
\begin{itemize}
	\item \emph{Part (i):} As long as the agents' gradient signs
	still disagree with each other by a non-negligible amount, the
	Laplacian sign-update is guaranteed to decrease the optimality residual
	$\overline{F}(k)$ by at least a fixed amount per iteration. Intuitively,
	whenever two connected agents disagree in gradient sign, at least one of them move toward the common optimum, and the network's spectral gap $\lambda_2$ converts this local disagreement into global progress.	
	\item \emph{Part (ii):} Because $\overline{F}(k)\ge 0$
	and each ``outside-the-threshold'' step removes a fixed amount of
	residual, this phase cannot last forever: it ends within
	$K_\varepsilon = O(1/\varepsilon^2)$ iterations, at which point the
	iterates have necessarily entered a small neighborhood of the
	invariant/optimal set.	
	\item \emph{Part (iii):} Once inside the
	threshold region, the sign information is too coarse to certify further descent which results in the chattering regime, and the iterations stay within an explicit, fixed-size ball around
	$x^\ast$, of radius $O(\tau)$, with the cost gap as $O(\tau^2)$.
\end{itemize}
}

\begin{rem}[Interpretation and the role of~$\tau$]\label{rem_threshold}
	{Theorem~\ref{thm_converg} clarifies the trade-off between step-size, convergence speed, and accuracy. The threshold $\eta(\tau) = O(\tau)$ defines a $\tau$-dependent neighborhood: outside this neighborhood the algorithm makes guaranteed progress; inside it, the coarse single-bit information is insufficient to ensure descent and the iterates may oscillate (the so-called chattering phenomena). The resulting optimality gap scales as $\overline{F} = O(\tau^2)$ and $\|\widetilde{\mb{x}}\|_2 = O(\tau)$.This yields a formal $\varepsilon$-convergence guarantee: for any prescribed accuracy $\varepsilon > 0$, choosing $\tau = O(\varepsilon)$ ensures the iterates reach a neighborhood with $\|\mb{x}(k)-\mb{x}^*\|_2 = O(\varepsilon)$ in finite time, at the cost of slower convergence ($K_\varepsilon = O(1/\varepsilon^2)$ iterations). This scaling is a fundamental limitation of fixed-step sign-based dynamics, not an artifact of the analysis.}
\end{rem}

\begin{rem}[Extension to $B$-connected time-varying graphs]\label{rem_Bconnected}
	When the graph is time-varying with Assumption~\ref{ass_G}(ii), the single-step displacement $\delta\mb{x}(\ell) = -\tau L_{\mc{G}}(\ell)\pmb{\phi}(\ell)$ depends on both the instantaneous graph and the current sign vector. Over a $B$-window, the cumulative displacement $\mb{x}(k{+}B) - \mb{x}(k) = -\tau\sum_{\ell=k}^{k+B-1}L_{\mc{G}}(\ell)\pmb{\phi}(\ell)$ generally differs from $-\tau\, L_{\mc{G}_B}(k)\pmb{\phi}(k)$ because $\pmb{\phi}(\ell)$ may change within the window.

	However, the change in $\pmb{\phi}(\ell)$ over the window is controlled: each step modifies $\mb{x}$ by at most $O(\tau)$ per component (see~\eqref{eq_epsil}), and changing any sign $\phi_i$ requires the corresponding gradient $\partial_{x_i}f_i$ to cross zero, which by strong convexity requires a state displacement bounded below by the local gradient margin. For $\tau$ sufficiently small (a sufficient condition depending on $\nu$, $B$, the edge weights, and the local gradient margins), the sign vector $\pmb{\phi}(\ell)$ remains constant over each $B$-window, and the analysis of Theorem~\ref{thm_converg} applies with $\lambda_2, \sigma_L$ replaced by the corresponding quantities $\lambda_{2,B}, \sigma_{L,B}$ of the union-graph Laplacian.
\end{rem}

\begin{rem}[Generality beyond the signum nonlinearity]\label{rem_general_convergence}
	The proof relies on two properties of $\pmb{\phi} = \mbox{sgn}(\nabla F)$: (i)~\emph{sign preservation} ($(\nabla F)_i\,\phi_i \geq 0$), ensuring the edge-sum in the descent term is non-negative; and (ii)~\emph{uniform boundedness} ($\|\pmb{\phi}\|_\infty \leq 1$), controlling the quadratic term. Any odd mapping $\varphi\colon\mathbb{R}\to\mathbb{R}$ with $u\,\varphi(u) > 0$ for $u \neq 0$ and $|\varphi(u)| \leq C$ shares these properties, with the threshold scaling as $O(C^2\tau)$. The signum function ($C=1$) minimizes communication to a single bit; power mappings and saturation functions are other members of this class that require multi-bit exchange but may yield tighter neighborhoods.
\end{rem}

The descent bound in Theorem~\ref{thm_converg} controls the optimality residual
$\overline{F}(k)$. We now translate this into a per-step bound on the error vector
$\widetilde{\mb{x}}(k) := \mb{x}(k) - \mb{x}^*$.

\begin{lem}[Error ratio bound outside the threshold region]\label{lem_error_rate}
Under the conditions of Theorem~\ref{thm_converg}, at every iteration~$k$ for which
$\pmb{\phi}(k) \notin \mathrm{span}(\mb{1}_n)$, $\|\nabla F(\mb{x}(k))\|_1 \geq \eta(\tau)$,
and $\overline{F}(k)>0$ (equivalently $\mb{x}(k)\neq \mb{x}^*$), the error
$\widetilde{\mb{x}}(k) := \mb{x}(k) - \mb{x}^*$ satisfies
\begin{align}\label{eq_xrate}
\frac{\|\widetilde{\mb{x}}(k+1)\|_2^2}{\|\widetilde{\mb{x}}(k)\|_2^2}
\;\leq\;
\frac{\mu}{\nu}\;\frac{\overline{F}(k+1)}{\overline{F}(k)}
\;\leq\;
\frac{\mu}{\nu}\,
\Bigl(
1 - \frac{1}{\overline{F}(k)}
\Bigl[\frac{\lambda_2\,\tau}{n}\,\|\nabla F(\mb{x}(k))\|_1 - \frac{\mu}{2}\,n\,\sigma_L^2\,\tau^2\Bigr]
\Bigr),
\end{align}
where the bracketed term is non-negative under the threshold condition
$\|\nabla F(\mb{x}(k))\|_1 \geq \eta(\tau)$.
\end{lem}

\begin{proof}
By the residual--error equivalence for $\nu$-strongly convex and $\mu$-smooth functions
(Lemma~\ref{lem_cortes}(ii)), we have
\begin{align}\label{eq_sandwich}
\frac{\nu}{2}\,\|\widetilde{\mb{x}}(k+1)\|_2^2
\;\leq\; \overline{F}(k+1),
\qquad
\overline{F}(k)
\;\leq\; \frac{\mu}{2}\,\|\widetilde{\mb{x}}(k)\|_2^2.
\end{align}
Since $\overline{F}(k)>0$ by assumption, dividing the first inequality by the second yields
\begin{align}\label{eq_ratio_sandwich}
\frac{\|\widetilde{\mb{x}}(k+1)\|_2^2}{\|\widetilde{\mb{x}}(k)\|_2^2}
\;\leq\;
\frac{\mu}{\nu}\;\frac{\overline{F}(k+1)}{\overline{F}(k)}.
\end{align}
From the combined bound~\eqref{eq_combined_bound} in the proof of Theorem~\ref{thm_converg},
\begin{align*}
\overline{F}(k+1)-\overline{F}(k)
\;\leq\;
-\frac{\lambda_2\,\tau}{n}\,\|\nabla F(\mb{x}(k))\|_1 \;+\; \frac{\mu}{2}\,n\,\sigma_L^2\,\tau^2,
\end{align*}
and dividing by $\overline{F}(k)>0$ gives
\begin{align}\label{eq_Fratio}
\frac{\overline{F}(k+1)}{\overline{F}(k)}
\;\leq\;
1 \;-\; \frac{1}{\overline{F}(k)}
\Bigl[\frac{\lambda_2\,\tau}{n}\,\|\nabla F(\mb{x}(k))\|_1 \;-\; \frac{\mu}{2}\,n\,\sigma_L^2\,\tau^2\Bigr].
\end{align}
When $\|\nabla F(\mb{x}(k))\|_1 \geq \eta(\tau)$, the bracketed term is non-negative
(cf.\ Step~4 of Theorem~\ref{thm_converg}), hence $\overline{F}(k+1)/\overline{F}(k)\le 1$.
Substituting~\eqref{eq_Fratio} into~\eqref{eq_ratio_sandwich} yields~\eqref{eq_xrate}.
\end{proof}

\begin{rem}[Interpretation of the error ratio bound]\label{rem_error_contraction}
The bound~\eqref{eq_xrate} separates the effect of conditioning (through $\mu/\nu$) from the
residual decay ratio $\overline{F}(k+1)/\overline{F}(k)\le 1$ ensured by Theorem~\ref{thm_converg}(i)
outside the threshold region.

Note that~\eqref{eq_xrate} does \emph{not} imply error contraction in general, because $\mu/\nu\ge 1$.
A sufficient condition for strict contraction at iteration~$k$ is
\[
\frac{\overline{F}(k+1)}{\overline{F}(k)} \;<\; \frac{\nu}{\mu},
\]
in which case $\|\widetilde{\mb{x}}(k+1)\|_2 < \|\widetilde{\mb{x}}(k)\|_2$.
This holds automatically for quadratic costs (where $\mu=\nu$), and more generally when
$\|\nabla F(\mb{x}(k))\|_1$ is sufficiently above $\eta(\tau)$ so that the residual decrease in
Theorem~\ref{thm_converg}(i) is large relative to $\overline{F}(k)$.

For ill-conditioned problems ($\mu/\nu\gg 1$), the bound allows the error norm to temporarily
increase even while $\overline{F}(k)$ decreases. This is a standard phenomenon in strongly convex
optimization and is not specific to the single-bit algorithm.
\end{rem}

\subsection{Chattering behavior and steady-state optimality gap}\label{sec_chattering}

Theorem~\ref{thm_converg} establishes that, under a fixed step-size~$\tau$, the iterates enter a neighborhood of the invariant set~$\mc{I}$ defined by the threshold $\eta(\tau) = O(\tau)$. Inside this neighborhood, the signum nonlinearity prevents the dynamics from settling at the exact optimizer~$\mb{x}^*$: the discrete sign vector $\pmb{\phi}(k)\in\{-1,0,+1\}^n$ cannot encode the vanishing gradient differences needed to steer $\mb{x}(k)$ toward $\mb{x}^*$, causing the iterates to oscillate -- a phenomenon known as \emph{chattering}. We now characterize the per-step oscillation amplitude and discuss its implications for the steady-state optimality gap.

\paragraph{Per-step displacement bound}
From the dynamics~\eqref{eq_sol}, the displacement of any single component satisfies
\begin{align}\label{eq_epsil}
	|x_i(k{+}1) - x_i(k)|
	\;=\;
	\tau\,\Bigl|\sum_{j \in \mc{N}^-_i(k)} W_{ij}(k)\bigl(\phi_i(k) - \phi_j(k)\bigr)\Bigr|
	\;\leq\;
	2\tau \sum_{j \in \mc{N}^-_i(k)} W_{ij}(k),
\end{align}
since $|\phi_i - \phi_j| \leq 2$. Aggregating over all nodes, we have
\begin{align}\label{eq_epsil1}
	\|\mb{x}(k{+}1) - \mb{x}(k)\|_1
	\;\leq\;
	2\tau\,\mc{W},
	\qquad
	\text{where}\quad \mc{W} := \sum_{i=1}^n \sum_{j \in \mc{N}^-_i} W_{ij}
\end{align}
is the total weighted out-degree of the network (equal to $\mb{1}_n^\top D\,\mb{1}_n$ for a static graph). Since $\|\mb{v}\|_2 \leq \|\mb{v}\|_1$ for any $\mb{v}\in\mathbb{R}^n$, the bound also holds in the $\ell_2$-norm.

The increment bound~\eqref{eq_epsil1} implies that in the chattering regime the iterates cannot move faster than $O(\tau)$ per step: any oscillations induced by sign switches have per-step amplitude at most $2\tau\mc{W}$. Note, however, that~\eqref{eq_epsil1} bounds only the \emph{step-to-step displacement}, not the distance $\|\mb{x}(k)-\mb{x}^*\|_2$ to the optimizer. The rigorous bound on the latter is provided by Theorem~\ref{thm_converg}(iii).

\paragraph{Two-cycle diameter bound}
If the iterates enter an eventual two-cycle (i.e., $\mb{x}(k{+}2) = \mb{x}(k)$ for all $k \geq k_0$), then~\eqref{eq_epsil1} directly bounds the cycle diameter:

\begin{lem}[Two-cycle diameter]\label{lem_twocycle}
	If the dynamics~\eqref{eq_sol} admits an eventual two-cycle $\{\mb{x}^a,\mb{x}^b\}$ with $\mb{x}(k) = \mb{x}^a$, $\mb{x}(k{+}1)=\mb{x}^b$, $\mb{x}(k{+}2) = \mb{x}^a$, then $\|\mb{x}^a - \mb{x}^b\|_2 \leq 2\tau\,\mc{W}$, and each point of the cycle lies within distance $\tau\mc{W}$ of the cycle midpoint $\tfrac{1}{2}(\mb{x}^a+\mb{x}^b)$.
\end{lem}
\begin{proof}
	The bound $\|\mb{x}^a - \mb{x}^b\|_2 = \|\mb{x}(k{+}1)-\mb{x}(k)\|_2 \leq 2\tau\mc{W}$ follows directly from~\eqref{eq_epsil1}. The midpoint claim follows by the triangle inequality: $\|\mb{x}^a - \tfrac{1}{2}(\mb{x}^a+\mb{x}^b)\|_2 = \tfrac{1}{2}\|\mb{x}^a-\mb{x}^b\|_2 \leq \tau\mc{W}$.
\end{proof}

Note that Lemma~\ref{lem_twocycle} bounds the diameter of the limit cycle, not the distance from the cycle to~$\mb{x}^*$. The latter requires the Lyapunov-based analysis and is provided by Theorem~\ref{thm_converg}(iii).

\paragraph{Steady-state optimality gap}
Combining these observations, the rigorous steady-state bounds are those of Theorem~\ref{thm_converg}(iii): for any $\varepsilon > 0$, once the iterates satisfy $\|\nabla F(\mb{x}(k))\|_1 \leq \eta(\tau)+\varepsilon$, we have
\begin{align}\label{eq_gap_summary}
	\|\mb{x}(k)-\mb{x}^*\|_2
	\;\leq\;
	\frac{1+1/\!\sqrt{n}}{\nu}\,\bigl(\eta(\tau)+\varepsilon\bigr),
	\qquad
	\overline{F}(k)
	\;\leq\;
	\frac{(1+1/\!\sqrt{n})^2}{2\nu}\,\bigl(\eta(\tau)+\varepsilon\bigr)^2.
\end{align}
Since $\eta(\tau) = O(\tau)$, both bounds vanish as $\tau\to 0$, confirming that the optimality gap can be made arbitrarily small by reducing the step-size.

{
\begin{rem}[Trade-off between convergence speed and accuracy]\label{rem_tradeoff}
	The step-size $\tau$ governs a fundamental trade-off: larger $\tau$ yields faster descent outside the threshold region (the per-step drop in~\eqref{eq_uniform_drop} scales as $O(\tau)$) but a wider optimality gap ($\eta(\tau) = O(\tau)$); smaller $\tau$ yields a tighter gap at the cost of slower convergence. Quantitatively, for a prescribed accuracy $\varepsilon > 0$, choosing $\tau = O(\lambda_2\varepsilon/(\mu n^2\sigma_L^2))$ ensures the gap~\eqref{eq_gap_summary} is $O(\varepsilon)$, and the first-entry time from Theorem~\ref{thm_converg}(ii) becomes $K_\varepsilon = O(\mu^2 n^3\sigma_L^2\overline{F}(0)/(\lambda_2^2\varepsilon^2))$ iterations. Hence, $\tau$ should be selected according to the desired accuracy rather than treated as an arbitrary parameter, i.e., not any arbitrary large $\tau$ is permissible given the prescribed accuracy $\varepsilon$ on threshold $\eta(\tau)$.
\end{rem}
}

{Next we examine how penalty terms affect the descent threshold and, consequently, the steady state optimality residual and residual error. 
Considering the quadratic penalty case, since $\mu_\sigma = \mu + 2\sigma$, we have
\begin{align}
	\eta(\tau, \sigma) = \frac{(\mu + 2\sigma)n^2 \sigma_L^2}{2\lambda_2}\tau.
\end{align}
Consequently, the steady-state neighborhood bounds from Theorem~1(iii) become
\begin{align}
	\|\mathbf{x}(k) - \mathbf{x}^*\|_2 \leq 
	\frac{1 + 1/\sqrt{n}}{\nu_\sigma} \left( \eta(\tau, \sigma) + \epsilon \right),
\end{align}
and
\begin{align}
	\overline{F}(k) \leq \frac{(1 + 1/\sqrt{n})^2}{2\nu_\sigma} 
	\left( \eta(\tau, \sigma) + \epsilon \right)^2.
\end{align}
These bounds explicitly quantify how increasing $\sigma$:
\begin{enumerate}
	\item \textit{Widens the steady-state neighborhood} linearly in $\sigma$ 
	through $\mu_\sigma$ in $\eta(\tau, \sigma)$;
	\item \textit{May require a smaller step-size} $\tau$ to maintain the 
	same accuracy, as the threshold scales with $\sigma$.
\end{enumerate}
The effect of $\sigma$ on chattering should be distinguished from its effect on conditioning. Our convergence analysis shows that, for a fixed admissible step size, the steady-state decision error induced by the one-bit sign operation is of order $O(\tau)$, while the corresponding cost residual is of order $O(\tau^2)$. Hence, increasing $\sigma$ does not by itself change the fundamental $O(\tau)$ scaling of the quantization-induced steady-state neighborhood. Instead, it changes the allowable range of $\tau$ through $\mu_\sigma$.
In practice, we recommend 
choosing $\sigma$ as the smallest value that achieves acceptable 
constraint violation, balancing accuracy versus convergence speed.
}

{Because only the sign of the gradient is transmitted, the algorithm does not retain the gradient magnitude needed to make the update vanish continuously as $\mb{x}(k)\rightarrow\mb{x}^*$. Consequently, for a fixed nonzero $\tau$, a nonzero steady-state neighborhood is expected (which is denoted by threshold $\eta(\tau)$). Our theorem does not claim exact convergence for fixed $\tau$; rather, it quantifies this neighborhood and proves that it shrinks to the optimizer as $\tau\rightarrow 0$.}

\paragraph{Diminishing step-sizes}
The chattering limitation is inherent to fixed-$\tau$ sign-based dynamics but can be mitigated by a diminishing step-size schedule. Replacing the fixed $\tau$ by $\tau(k) = a/k$ for some $a > 0$, the per-step displacement~\eqref{eq_epsil1} decays as $O(1/k)$, so the oscillation amplitude vanishes over time. Since $\sum_k \tau(k) = \infty$ and $\sum_k \tau(k)^2 < \infty$ (the Robbins--Monro conditions), one expects the iterates to converge to $\mb{x}^*$ asymptotically. However, the discontinuity of the signum map requires care: standard stochastic-approximation or subgradient convergence results (cf.~\cite{nesterov1998introductory}) do not apply directly to sign-based dynamics without additional regularity assumptions (e.g., excluding persistent sign switches at the boundary $\partial_{x_i}f_i = 0$). A formal convergence rate analysis under diminishing step-sizes is an interesting direction for future work; the simulations in Section~\ref{sec_sim} confirm that diminishing $\tau$ yields lower residuals than a fixed step-size (see Fig.~\ref{fig_sim1}).

\section{Simulations} \label{sec_sim}

We verify Algorithm~\ref{alg_1} on three test-beds of increasing scale.
Every experiment is run in MATLAB R2023b on a laptop with an
Intel Core~i5 at 3.30\,GHz, 8\,GB RAM, and 512\,GB SSD.
Throughout, we report the \emph{optimality residual}
      $\overline{F}(k):=F(\mb{x}(k))-F(\mb{x}^*)$, which
      Theorem~\ref{thm_converg} guarantees to decrease
      monotonically outside the invariant set~$\mc{I}$.

\subsection{Step-Size Scenarios and Comparison with the Literature}\label{sec:sim:stepsize}

\paragraph{Setup}
We consider $n=8$ agents over a sparse time-varying network
composed of four directed subgraphs $\mc{G}_1,\ldots,\mc{G}_4$,
none of which is individually connected.
Graphs $\mc{G}_1,\mc{G}_2$ form a directed cycle with unit edge
weights; $\mc{G}_3,\mc{G}_4$ together form a 3-hop digraph with
edge weights equal to~2.
The network switches cyclically every 10~iterations, so the
union over any window of $B=40$ consecutive steps is
weight-balanced and strongly connected, satisfying
Assumption~\ref{ass_G}.
The cost functions follow~\cite{doan2017scl}:
$f_i(x_i)=\alpha_i\bigl((x_i-\beta_i)^4+(x_i-\beta_i)^2\bigr)$
with $\alpha_i\in[0,2]$, $\beta_i\in[0,5]$ drawn uniformly at
random, and the demand constraint is $\sum_i x_i=b=20$.

\begin{figure}[t]
	\centering
	\includegraphics[width=2.65in]{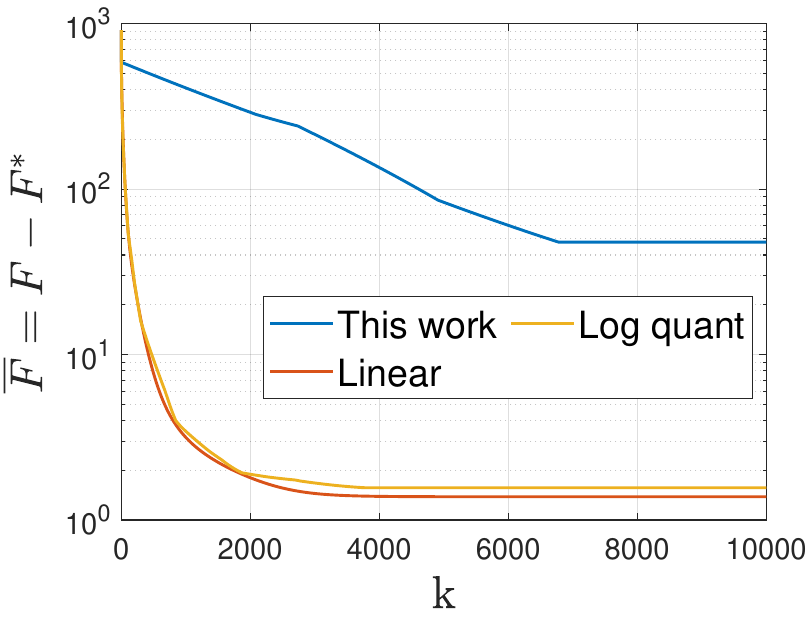}
	\includegraphics[width=2.65in]{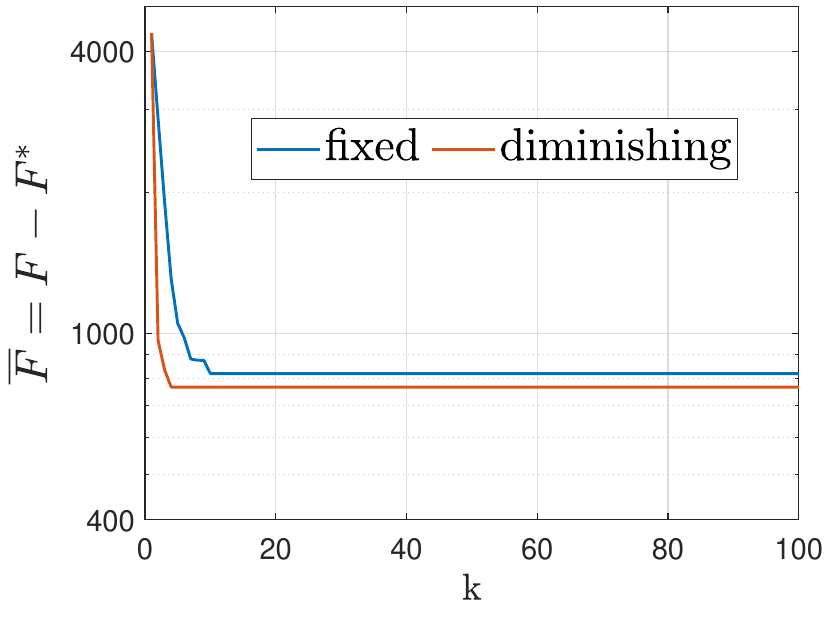}
	\caption{(Left) Optimality residual $\overline{F}(k)$
	for the single-bit algorithm versus the linear
	method~\cite{doan2017scl} and log-quantized
	method~\cite{my_cpu}.  The single-bit residual floor is
	consistent with Theorem~\ref{thm_converg}(iii).
	(Right) Fixed step-size $\tau=0.01$ versus diminishing
	$\tau(k)=0.05/k$; the diminishing schedule drives the
	residual below the fixed-$\tau$ floor.}
	\label{fig_sim1}
\end{figure}

\paragraph{Fixed step-size comparison}
We set $\tau=0.0001$ and compare Algorithm~\ref{alg_1} against
two baselines: the linear (full-gradient) method
of~\cite{doan2017scl} and the log-quantized method
of~\cite{my_cpu}.
Figure~\ref{fig_sim1} (top) plots $\overline{F}(k)$ versus
iteration~$k$.
The single-bit algorithm converges to a neighborhood of
$\mb{x}^*$ whose size is consistent with the bound in
Theorem~\ref{thm_converg}(iii), while the two multi-bit methods
converge to the exact optimum.
This confirms the theoretical trade-off: the residual floor is
the price of reducing communication from 64 (or
$\lceil\log_2\rceil$) bits per link to a single bit. The comutational cost is compared in Table~\ref{tab_comp00}. {For fair comparison, the running time is reported over the same number of $k_{\max}{=}10000$ iterations for all three algorithms. }

\paragraph{Communication cost comparison}
{The proposed method consists of two communication phases. First, a one-time initialization routine is used to construct a feasible starting point. This routine may require multi-bit message passing over a spanning tree. This initialization cost is incurred only once. Second, during the iterative phase,
each iteration of the linear method~\cite{doan2017scl} transmits
64~bits per link; the log-quantized method~\cite{my_cpu} uses
5~bits (quantization level~$\delta=1$); Algorithm~\ref{alg_1}
uses 1~bit.
Despite requiring more iterations, the single-bit algorithm
reaches the target with fewer total bits than
the log-quantized and the linear
method.}

\begin{table}[t]
\centering
\caption{{Comparison on the $n{=}8$ time-varying network over $10000$ iterations ($\tau{=}0.0001$).}} \label{tab_comp00}
\smallskip
{
\begin{tabular}{lccc}
\hline
Algorithm & Run time& Optimization & Initialization\\
\hline
Single-bit (Alg.~\ref{alg_1}) & 0.0661\,s & 0.0614\,s & 0.0047 \,s\\
Log-quantized~\cite{my_cpu} & 0.1981\,s & 0.1934\,s& 0.0047 \,s\\
Linear~\cite{doan2017scl} & 0.0640\,s & 0.0593\,s & 0.0047 \,s\\
\hline
\end{tabular}
}
\end{table}

\paragraph{Diminishing step-size}
Figure~\ref{fig_sim1} (bottom) compares fixed $\tau=0.01$ with
the diminishing schedule $\tau(k)=0.05/k$.
The diminishing schedule satisfies the Robbins--Monro conditions
($\sum_k\tau(k)=\infty$, $\sum_k\tau(k)^2<\infty$) and drives
the residual below the fixed-$\tau$ floor, illustrating that
convergence closer to $\mb{x}^*$ is attainable at the expense of
a slower rate.
The bound in Theorem~\ref{thm_converg}(iii) applies at each~$k$
with the instantaneous $\tau(k)$: the shrinking gap
$\overline{F}=O(\tau(k)^2)=O(1/k^2)$ explains the monotonically
decreasing residual envelope.

\subsection{Application to Energy Management and Comparison
            with the Literature}\label{sec:sim:edp}

\paragraph{Setup}
We consider $n=1000$ power generators connected over a
weight-balanced Erd\H{o}s--R\'enyi digraph with average
out-degree~10.
The quadratic cost model is
$f_i(x_i)=a_ix_i^2+c_ix_i$ with
$a_i\in\{0.03,0.035,0.04\}$,
$c_i\in\{2,2.5,3,4\}$, following~\cite{kar_edp,kar2012distributed}.
The aggregate demand is $b=60\,000$\,MW.
Local generation limits $\underline{m}=20$\,MW and
$\overline{m}\in\{70,80,90\}$\,MW (depending on generator type)
are enforced via the quadratic penalty
$f_i^\sigma(x_i)=f_i(x_i)+h^2(x_i-\overline{m})+h^2(\underline{m}-x_i)$
with $\sigma=50$; under this choice the penalized
strong-convexity constant is
$\nu_\sigma=2\min_i a_i=0.06$ and the smoothness constant is
$\mu_\sigma=2\max_i a_i+2\sigma=100.08$, so
Assumption~\ref{ass_strong} holds with these modified constants
and all bounds in Theorem~\ref{thm_converg} apply accordingly.

\begin{figure}[t]
	\centering
	\includegraphics[width=2.65in]{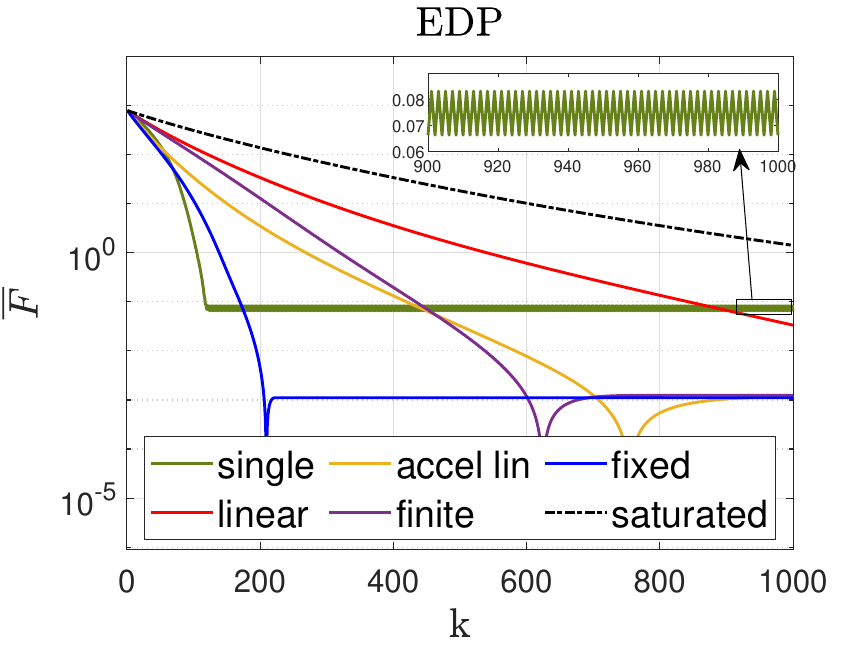}
	\includegraphics[width=2.65in]{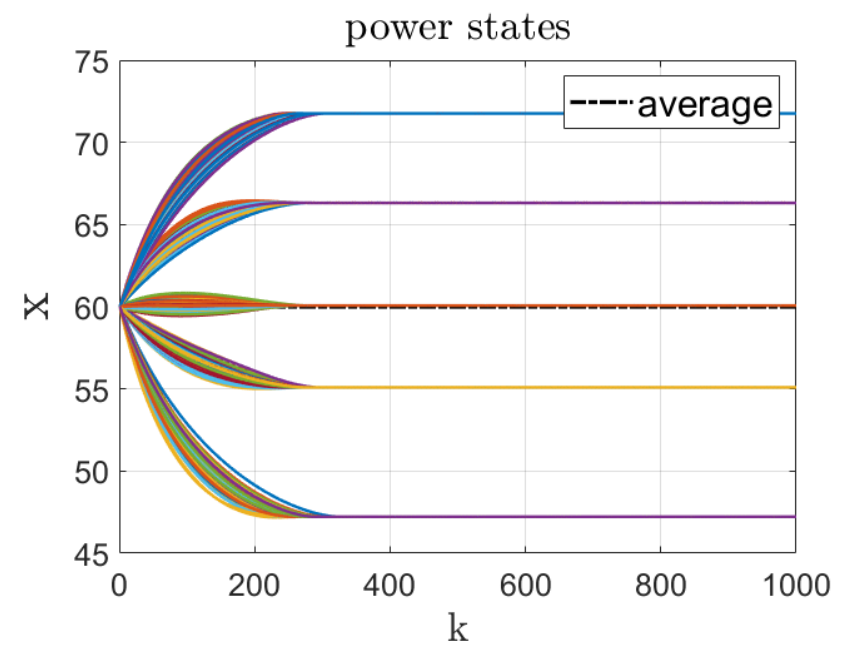}
	\caption{(Left) Optimality residual for the EDP:
	Algorithm~\ref{alg_1} versus five multi-bit baselines.
	(Right) Power states $x_i(k)$ at all generators and
	their average $\frac{1}{n}\sum_i x_i(k)=b/n$,
	confirming anytime feasibility
	(Lemma~\ref{lem_feasible_intime}).}
	\label{fig_edp}
\end{figure}

\paragraph{Residual and feasibility}
Figure~\ref{fig_edp} (left) plots $\overline{F}(k)$ for
Algorithm~\ref{alg_1} ($\tau=0.1$) alongside five multi-bit
baselines: the linear method~\cite{boyd2006optimal}, the
accelerated (heavy-ball) method~\cite{shames2011accelerated},
saturated dynamics~\cite{ccta22}, and the
finite-time~\cite{chen2016distributed} and
fixed-time~\cite{spl} nonlinear schemes.
Algorithm~\ref{alg_1} converges to a neighborhood consistent
with the $O(\tau^2)$ residual bound in
Eq.~\eqref{eq_gap_summary}, while all multi-bit methods reach
machine precision.

Figure~\ref{fig_edp} (right) shows the individual power states
$x_i(k)$ and their average
$\frac{1}{n}\sum_i x_i(k)=b/n$.
The average remains constant at every iteration, verifying the
anytime feasibility guarantee of Lemma~\ref{lem_feasible_intime}.

\paragraph{Communication and computation cost}
Table~\ref{tab_comp0} reports wall-clock run times and the
per-link bit-rate. {For fair comparison, the running time is reported over the same number of $k_{\max}{=}1000$ iterations for all algorithms. }
Because every baseline transmits 64~bits per link per iteration
while Algorithm~\ref{alg_1} transmits~1, the single-bit method
incurs the lowest total bit-rate despite requiring more
iterations to reach comparable residual levels.
Its per-iteration computational cost is also among the smallest, since
the signum evaluation replaces floating-point multiplications.

\begin{table}[t]
\centering
\caption{{Economic dispatch ($n{=}1000$, $\tau{=}0.1$,
$k_{\max}{=}1000$ iterations).}}\label{tab_comp0}
\smallskip
{
\begin{tabular}{lccc}
\hline
Algorithm & Run time & Optimization & Initialization\\
\hline
Single-bit (Alg.~\ref{alg_1})       &  11.26\,s & 10.67\,s & 0.59\,s \\
Linear~\cite{boyd2006optimal}        & 11.07\,s & 10.48\,s &  0.59\,s \\
Accelerated~\cite{shames2011accelerated} & 11.41\,s & 10.82\,s &0.59\,s\\
Finite-time~\cite{chen2016distributed}   & 24.96\,s & 24.37\,s &0.59\,s \\
Fixed-time~\cite{spl}                & 38.39\,s & 37.80\,s &0.59\,s \\
Saturated~\cite{ccta22}              & 11.58\,s & 10.99 \,s &0.59\,s  \\
\hline
\end{tabular}}
\end{table}

\begin{rem}\label{rem:penalty}
The penalty weight $\sigma=50$ introduces a trade-off:
increasing $\sigma$ tightens the box-constraint approximation
but raises $\mu_\sigma$ and, through the descent threshold
$\eta(\tau)$ in Theorem~\ref{thm_converg}, may require
a smaller~$\tau$.
In this experiment the resulting neighborhood satisfies the
generation limits to within $\pm 0.3$\,MW, which is well within
operational tolerance.
\end{rem}

\subsection{Application to CPU Scheduling and Comparison
            with the Literature}\label{sec:sim:cpu}

\paragraph{Setup}
A network of $n=500$ servers is arranged in a cyclic
(weight-balanced, strongly connected) digraph.
Each server $i$ has cost function~\cite{rikos2021optimal,my_cpu}:
\begin{align}\label{eq:fiz2}
	f_i(x_i) = \frac{1}{2\pi_i^{\max}} (x_i - \rho_i)^2
\end{align}
where $x_i,\rho_i$ respectively denote the allocated CPU
workload and positive demand at node (server) $i$ and
$\sum_{i=1}^n x_i = \rho$ with $\rho$ denoting the overall CPU
demands.
The parameter $\pi_i^{\max}$ is defined as
$\pi_i^{\max} \coloneqq c_i T_h$, where $c_i$ is the sum of
all clock rate frequencies of all processing cores of
node/server $i$ given in cycles/second and $T_h$ is the CPU
allocation time period.
For this simulation we set $\pi_i^{\max}=80$,
$\rho_i \in [10~30]$, and $\rho = 10000$.
The local box constraints on the allocated CPUs are
$m_i = 0 \leq x_i \leq 0.75 \pi_i^{\max} = M_i$, which are
addressed via logarithmic penalty functions with $\kappa=2$.
The network of servers is considered as a simple cyclic graph.
In Algorithm~\ref{alg_1} we set $\tau = 1$.

\begin{figure}[t]
	\centering
	\includegraphics[width=2.65in]{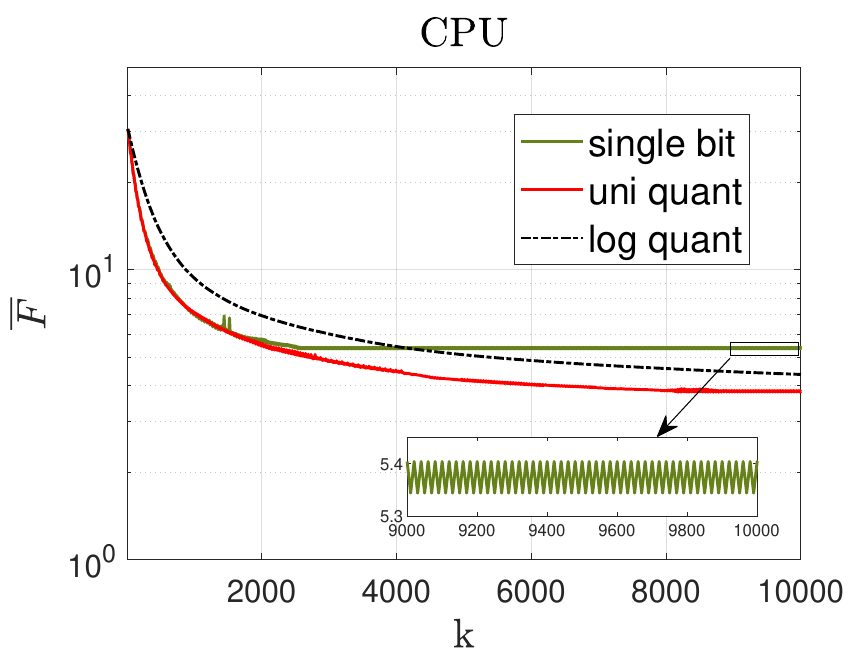}
	\includegraphics[width=2.65in]{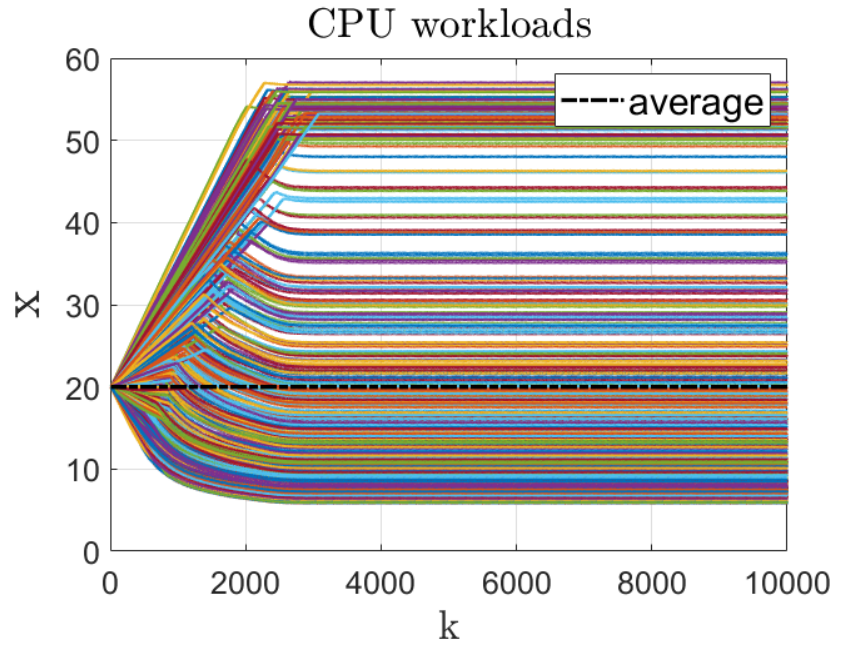}
	\caption{(Left) Optimality residual for CPU scheduling:
	Algorithm~\ref{alg_1} versus log-quantized~\cite{my_cpu}
	and uniformly-quantized~\cite{rikos2021optimal} methods.
	(Right) Allocated workloads $x_i(k)$ at all servers and
	their average, confirming anytime feasibility
	(Lemma~\ref{lem_feasible_intime}).}
	\label{fig_cpu}
\end{figure}

\paragraph{Results}
Figure~\ref{fig_cpu} (left) compares Algorithm~\ref{alg_1} with
the log-quantized~\cite{my_cpu} and
uniformly-quantized~\cite{rikos2021optimal} schedulers
(quantization level $\delta=1/32$ in both cases).
The single-bit method converges to a slightly wider
neighborhood, consistent with Theorem~\ref{thm_converg}(iii),
but at substantially lower communication cost: 1~bit versus
5~bits (log) or 6~bits (uniform) per link.

Figure~\ref{fig_cpu} (right) confirms anytime feasibility
(Lemma~\ref{lem_feasible_intime}): the total allocated workload
$\sum_i x_i(k) = \rho$ is preserved at every iteration.

\begin{table}[t]
\centering
\caption{{CPU scheduling ($n{=}500$, $\tau{=}1$,
$k_{\max}{=}10\,000$).  Comparison of the computational
complexity for the cost given by~\eqref{eq:fiz2}.}}\label{tab_comp}
\smallskip
{
\begin{tabular}{lccc}
\hline
Algorithm & Run time & Initialization & Optimization\\
\hline
Single-bit (Alg.~\ref{alg_1})
  & 5.81\,s & 0.21\,s & 5.60\,s \\
Log-quant.~\cite{my_cpu}
  & 21.58\,s& 0.21\,s & 21.37\,s \\
Uniform-quant.~\cite{rikos2021optimal}
  & 8.86\,s  & 0.21\,s & 8.65\,s\\
\hline
\end{tabular}
}
\end{table}

\begin{rem}\label{rem:cpu}
The proposed single-bit algorithm for CPU scheduling advances
the existing literature in two respects:
(i)~it converges under single-bit data exchange, in contrast to
the multi-bit transfer required
by~\cite{my_cpu,rikos2021optimal}, significantly reducing the
communication load on the network; and
(ii)~its per-iteration run time is the lowest among all tested
methods (Table~\ref{tab_comp}), implying reduced computational
load on the servers.
\end{rem}

\paragraph{Summary of empirical findings}
Across all three experiments the simulations corroborate the
main theoretical predictions:
(i)~anytime feasibility (Lemma~\ref{lem_feasible_intime}) holds
exactly at every iteration;
(ii)~the optimality residual enters and remains within the
$O(\tau^2)$ neighborhood predicted by
Theorem~\ref{thm_converg}(iii); and
(iii)~diminishing step-sizes drive the residual below the
fixed-$\tau$ floor.

\begin{rem}
{Note that the optimization variables in our framework should not necessarily be interpreted as instantaneous actuator commands. In a practical implementation, the optimization layer would normally provide setpoints or references to a lower-level controller that accounts for actuator dynamics and other physical constraints. Consequently, the bounded oscillations observed in the optimization iterates do not automatically imply that physical actuators must switch at the same frequency or with the same amplitude.	However, in applications where the optimization output is closely applied to physical devices, for example in economic dispatch, rapid variations of generator setpoints could lead to undesirable interactions with generator governor dynamics if the optimization update rate is comparable to the actuator bandwidth. Similarly, in CPU scheduling, rapid oscillations in the requested CPU frequency could result in excessive dynamic voltage/frequency transitions or undesirable power fluctuations. These effects are implementation-dependent and are therefore not captured by the present optimization model. To mitigate such effects in practical cases, the optimization layer can operate at a slower rate than the low-level actuator controller, or the computed setpoints can be passed through an actuator-aware tracking/rate-limiting setup.}
\end{rem}

\section{Conclusions and Future Work} \label{sec_conclusion}

We have presented a single-bit distributed resource allocation algorithm based on Laplacian-gradient dynamics, in which each agent broadcasts only the sign of its local cost gradient. The Laplacian structure on weight-balanced digraphs preserves the demand constraint at every iterate, enabling anytime termination -- unlike ADMM or primal-dual methods that achieve feasibility only asymptotically.

Under a fixed step-size and uniform connectivity of the union graph, the iterates enter a neighborhood of the optimizer in finite time, with the optimality gap scaling linearly in the step-size (decision variable) and quadratically (cost residual). A formal $\varepsilon$-convergence guarantee follows directly: any prescribed accuracy is attained by choosing a proportionally small step-size. The analysis relies only on sign-preservation and uniform boundedness, so the single-bit protocol is the communication-minimizing instance of a broader class of odd nonlinearities sharing the same Lyapunov structure.

Several limitations remain open. The feasible initialization may require a one-time multi-bit spanning-tree routine, partially offsetting the single-bit savings in short-horizon settings. The penalty-function treatment of box constraints provides only approximate satisfaction, with the constraint tightness, step-size, and penalty weight coupled in a trade-off that deserves sharper characterization. Finally, while diminishing step-sizes eliminate the residual gap empirically, a formal convergence rate analysis is complicated by the discontinuity of the signum at zero and remains an important direction for future work, alongside noise resilience, connections to the compressed-gradient literature, and extensions to non-balanced digraphs and multi-dimensional decisions. {Moreover, communication delays and asynchronous packet arrivals are not analyzed in the present work and constitute an interesting direction for future research.}

{The underlying approach can be extended to the vector case. 
	Specifically, suppose that ($x_i\in\mathbb{R}^d$). The sign operation can be applied entry-wise, i.e., the transmitted message consists of the signs of the $d$ coordinates. With this modification, the key properties used in our analysis continue to hold. In particular, feasibility is preserved, the KKT characterization remains valid, and the convergence argument extends to the vector setting. The only change in the convergence bound is in the graph-dependent constant, which acquires the corresponding dimension-dependent factor (of order $d$) due to the norm conversion between the coordinate-wise and vector representations. This extension is an important direction for future research.}

\bibliographystyle{elsarticle-num}
\bibliography{bibliography}

\end{document}